\documentclass[iop]{emulateapj}

\newcommand\pcc{\;{\rm cm}^{-3}}
\newcommand\Msun{\; {\rm M}_{\odot}}
\newcommand\kms{\; {\rm km}\;{\rm s}^{-1}}
\newcommand\ergs{\; {\rm erg}\;{\rm s}^{-1}}

\newcommand\cm{\;{\rm cm}}
\newcommand\yr{\; {\rm yr}}
\newcommand\Myr{\;{\rm Myr}}

\newcommand\pc{\;{\rm pc}}
\newcommand\kpc{\;{\rm kpc}}
\newcommand\sfrunit{\Msun \kpc^{-2} \yr^{-1}}
\newcommand\Punit{\pcc\,{\rm K}}
\newcommand\Surf{\Msun\;{\rm pc^{-2}}}
\newcommand\rhounit{\Msun\;{\rm pc^{-3}}}
\newcommand\Kel{\;{\rm K}}
\newcommand\eV{\;{\rm eV}}

\newcommand\simgt{\lower.5ex\hbox{$\; \buildrel > \over \sim \;$}}
\newcommand\simlt{\lower.5ex\hbox{$\; \buildrel < \over \sim \;$}}
\newcommand\pderiv[2]{\frac{\partial {#1}}{\partial {#2}}}

\newcommand\advect[2][\vel]{{#1}\cdot \nabla {#2}}
\newcommand\rbrackets[1]{\left({#1}\right)}
\newcommand\sbrackets[1]{\left[{#1}\right]}
\newcommand\cbrackets[1]{\left\{{#1}\right\}}
\newcommand\abrackets[1]{\left\langle{#1}\right\rangle}
\newcommand\divergence[2][\rbrackets]{\nabla \cdot #1{#2}}

\newcommand\vel{\mathbf{v}}
\newcommand\Om{\mathbf{\Omega}}

\newcommand\xhat{\hat{\mathbf{x}} }
\newcommand\yhat{\hat{\mathbf{y}} }
\newcommand\zhat{\hat{\mathbf{z}} }

\newcommand\kbol{k_{\rm B}}

\newcommand\eff{\epsilon_{\rm ff}}
\newcommand\tff{t_{\rm ff}}

\newcommand\edyn{\epsilon_{\rm dyn}}
\newcommand\ever{\epsilon_{\rm ver}}
\newcommand\tver{t_{\rm ver}}
\newcommand\tdyn{t_{\rm dyn}}
\newcommand\Msn{m_{\rm *}}
\newcommand\rsh{r_{\rm sh}}

\newcommand\Psn{p_{\rm *}}

\newcommand\rhoth{\rho_{\rm cr}}
\newcommand\Tth{T_{\rm cr}}

\newcommand\nth{n_{\rm cr}}
\newcommand\SigSFR{\Sigma_{\rm SFR}}

\newcommand\SigSFRsun{\Sigma_{\rm SFR,0}}

\newcommand\Sigdiff{\Sigma_{\rm diff}}
\newcommand\torb{t_{\rm orb}}
\newcommand\tbin{t_{\rm bin}}
\newcommand\Pmax{P_{\rm max}}
\newcommand\Pmin{P_{\rm min}}

\newcommand\rhosd{\rho_{\rm sd}}
\newcommand\ngbc{n_{\rm GBC}}
\newcommand\fdiff{f_{\rm diff}}

\newcommand\sz{\sigma_{\rm z}}
\newcommand\szdiff{\sigma_{\rm z,diff}}
\newcommand\vzdiff{v_{\rm z,diff}}
\newcommand\vthdiff{v_{\rm th,diff}}

\newcommand\Hdiff{H_{\rm diff}}
\newcommand\Pth{P_{\rm th}}
\newcommand\Pturb{P_{\rm turb}}

\newcommand\Ptot{P_{\rm tot}}
\newcommand\rhomid{n_0}

\newcommand\PtotDE{P_{\rm tot, DE}}
\newcommand\Ptwo{P_{\rm two}}

\newcommand\frad{f_{\rm rad}}

\newcommand\etath{\eta_{\rm th}}
\newcommand\etaturb{\eta_{\rm turb}}

\newcommand\Pdriv{P_{\rm driv}}

\shorttitle{Regulation of Star Formation Rates in 3D}
\shortauthors{KIM, OSTRIKER, \& KIM}

\begin{document}

\title{Three Dimensional Hydrodynamic Simulations of Multiphase Galactic Disks
with Star Formation Feedback: I. Regulation of Star Formation Rates}

\author{Chang-Goo Kim\altaffilmark{1}, Eve C.\ Ostriker\altaffilmark{2,3}, and
Woong-Tae Kim\altaffilmark{4}}

\affil{$^1$Department of Physics \& Astronomy, University of Western
Ontario, London, Ontario N6A 3K7, Canada}
\affil{$^2$Department of
Astronomy, University of Maryland, College Park, MD 20742, USA}
\affil{$^3$Department of Astrophysical Sciences, Princeton
University, Princeton, NJ, 08544, USA}
\affil{$^4$Center for the
Exploration of the Origin of the Universe (CEOU), Astronomy Program,
Department of Physics \& Astronomy, Seoul National University, Seoul
151-742, Republic of Korea}
\email{ckim256@uwo.ca,
eco@astro.princeton.edu, wkim@astro.snu.ac.kr}

\begin{abstract}
The energy and momentum feedback from young stars has a profound
impact on the interstellar medium (ISM), including heating and driving
turbulence in the neutral gas that fuels future star formation.
Recent theory has argued that this leads to a quasi-equilibrium
self-regulated state, and for outer atomic-dominated disks results in
the surface density of star formation $\SigSFR$ varying approximately
linearly with the weight of the ISM (or midplane turbulent + thermal
pressure).  We use three-dimensional numerical hydrodynamic
simulations to test the theoretical predictions for thermal,
turbulent, and vertical dynamical equilibrium, and the implied
functional dependence of $\SigSFR$ on local disk properties.  Our
models demonstrate that all equilibria are established rapidly, and
that the expected proportionalities between mean thermal and turbulent
pressures and $\SigSFR$ apply.  For outer disk regions, this results
in $\SigSFR \propto \Sigma \sqrt{\rhosd}$, where $\Sigma$ is the total
gas surface density and $\rhosd$ is the midplane density of the
stellar disk (plus dark matter).  This scaling law arises because
$\rhosd$ sets the vertical dynamical time in our models (and outer
disk regions generally).  The coefficient in the star formation law
varies inversely with the specific energy and momentum yield from
massive stars.  We find proportions of warm and cold atomic gas,
turbulent-to-thermal pressure, and mean velocity dispersions that are
consistent with Solar-neighborhood and other outer-disk observations.
This study confirms the conclusions of a previous set of simulations,
which incorporated the same physics treatment but was restricted to
radial-vertical slices through the ISM.
\end{abstract}

\keywords{galaxies: ISM --- galaxies: kinematics and
dynamics --- galaxies: star formation --- method: numerical ---
turbulence}

\section{Introduction}\label{sec:intro}

Disk galaxies like the Milky Way are long-lived systems, evolving (in the
absence of interactions) only on timescales of several orbits.  The
interstellar medium (ISM) also evolves slowly overall, such that there are
well-defined ISM properties when averaged over a spatial domain of several disk
scale heights and several tenths of an orbit.  On small spatial scales,
however, the ISM is highly structured, and can change rapidly as both thermal
and dynamical timescales are short.  These short timescales suggest that the
ISM will be able to evolve to a quasi-equilibrium state, in which heating
balances cooling, and in which the mean pressure in the diffuse gas at any
height balances the weight of the overlying ISM.  The ideas that thermal and
vertical ``hydrostatic'' equilibrium should approximately apply are widely held
\citep[see e.g. the reviews of ][]{2001RvMP...73.1031F,2005ARA&A..43..337C} and
have been studied in detail over many decades in the astrophysical literature.
Turbulence has only been studied more recently with the advent of numerical
hydrodynamic simulations, but it too is expected to reach an equilibrium
between driving and dissipation \citep[cf.][]{sto98,1999ApJ...524..169M}.
Although thermal pressure plays a role, turbulent pressure is believed to be
the most important contributor to vertical force balance in the atomic ISM
\citep{1991ApJ...382..182L}.

In models of the atomic ISM, the heating rate per particle, and therefore the
equilibrium thermal pressure, is generally treated as an independent parameter,
set based on empirical values of ambient UV, X-rays, and cosmic rays, with
far-UV from young stars the dominant term \citep[e.g.][]{wol95,wol03}.  Many
processes may contribute to turbulent driving
\citep[e.g.][]{2004ARA&A..42..211E}, but those associated with feedback from
young stars are likely to be the most important on scales relevant for vertical
dynamical equilibrium within the disk.  Similar to the case for thermal
pressure, empirical measures of the supernova rate have often been used to
obtain predictions for the turbulent pressure in ISM models
\citep[e.g.][]{mck77}.

For a self-consistent ISM model, the equilibrium pressure obtained by balancing
various gain and loss terms must be the same as the equilibrium
pressure that offsets the vertical weight of the gas.  \citet[][hereafter
OML10]{2010ApJ...721..975O} and \citet[][hereafter OS11]{2011ApJ...731...41O}
used this principle, in combination with relations between turbulent and
thermal driving and the star formation rate, to obtain predictions for the
equilibrium star formation rate in disk systems regulated by feedback.  In this
model framework, the radiation fields and supernova rates that control
thermal and turbulent pressure in the ISM are no longer considered independent
(or empirically-determined) parameters, but must evolve (together with an
evolving star formation rate) to levels that yield pressures that are
consistent with vertical dynamical equilibrium.

\citet[][hereafter Paper~I]{2011ApJ...743...25K} tested this feedback-regulated
simultaneous equilibrium model via numerical hydrodynamic simulations for the
outer-disk regime in which gas is primarily atomic (both warm and cold), while
\citet[][hereafter SO12]{2012ApJ...754....2S} tested the equilibrium model with
simulations of the starburst regime in which gas is mainly molecular and cold.
Both of these numerical studies demonstrated that models for a wide range of
parameters indeed evolve to a quasi-steady, turbulent state.  Key physical
quantities, such as the star formation rate surface density ($\SigSFR$), disk
scale height, midplane thermal and turbulent pressures, warm and cold gas
fractions, and mass-weighted velocity dispersions, reach well-defined mean
values within an orbital time.  The saturation of these statistical properties
enables a comparison with the theory of self-regulated star formation developed
in OML10 and OS11, and also provides calibrations of certain parameters that
enter the theory.

In this paper, we return to the atomic-dominated outer disk regime, but extend
the simulations of Paper~I from two-dimensional radial-vertical domains to
fully three-dimensional models.  We explore a wide range of total gas surface
density $\Sigma\sim2.5 - 20\Surf$ and midplane density of stars plus dark
matter $\rhosd\sim 0.003 - 0.45\rhounit$.  We also include sheared galactic
rotation, focusing on model families in which the angular speed $\Omega
\propto\Sigma$ so that the Toomre parameter in the gas is constant ($Q \sim 2$
for saturated-state turbulence velocity dispersion $\sim7\kms$).  Our feedback
prescription includes both time-dependent heating and turbulent driving
dependent on the star formation rate, as described in Section \ref{sec:eq}. We
also vary the heating efficiency $\frad\sim0.2-5$ that connects the heating
rate and the SFR surface density.  This allows us to explore how e.g. varying
dust abundance (which would alter radiation penetration and heating) may affect
the saturated state and evolution.

As we shall describe, our simulations show that realistic star formation rates
are obtained when momentum feedback at the levels expected from the
corresponding Type II supernova rates are included.  This affirms the
conclusions reached in our previous numerical studies (Paper~I, SO12).  Although
detailed prescriptions differ, recent work from other groups has reached
similar conclusions regarding the ability of sufficient momentum feedback to
self-regulate star formation at realistic levels
\citep[e.g.][]{dob11,hop11,age13}.  Beyond simply demonstrating
that feedback is effective in self-regulating star formation, we also show
(following Paper~I) that star formation rates obey a near-linear scaling with
the pressure of the diffuse ISM, and we relate the coefficient to the inverse
of the specific momentum injected by massive stars.   This explains empirical
correlations of molecular gas and star formation with pressure identified by
\citet{bli04,bli06} and \citet{ler08}.  Our numerical results confirm the
conclusion of Paper~I that both warm and cold atomic gas are expected to be
present in the ISM for a wide range of conditions, consistent with observations
of both the Solar neighborhood \citep{hei03} and distant outer disk regions
\citep{2009ApJ...693.1250D,2013arXiv1304.7770P} of the Milky Way.

The plan of this paper is as follows.  In Section \ref{sec:method}, we
summarize our numerical methods and the parameter sets chosen for our
simulations modeling varying galactic environments.  In Section
\ref{sec:result} we present our results, including an overview of evolution
(Section \ref{sec:evol}), statistics of saturated-state properties for
different models (Section \ref{sec:stat}),  comparison with the predictions for
vertical dynamical, thermal, and turbulent equilibrium based on these
statistics (Section \ref{sec:equil}), and comparison with the theoretical
predictions for equilibrium star formation scalings and rates (Section
\ref{sec:scaling}).  We summarize and discuss our main conclusions in Section
\ref{sec:summary}.

\section{Numerical Methods and Models}\label{sec:method}
\subsection{Numerical Methods}\label{sec:eq}

We consider local ``shearing box'' models of galactic disks in  three
dimensions \citep[e.g.][]{kim02a,kim03}.  The axes in the local Cartesian frame
are $x\equiv R-R_0$ and $y\equiv R_0(\phi-\Omega t)$, where $R_0$ is the
galactocentric radius at the center of the domain and $\Omega\equiv\Omega(R_0)$
is the angular velocity at $R_0$;  $z$ is the  vertical coordinate centered on
the galactic midplane.  The background velocity in this local frame relative to
the domain center (at $x=y=z=0$) has the form $\vel_0=-q\Omega x\yhat$, where
$q\equiv -(d\ln \Omega/d\ln R)|_{R_0}$ is the local shear rate.  We assume a
flat rotation curve, such that we set $q=1$ and the local epicyclic frequency
is $\kappa=\sqrt{2}\Omega$.  The shearing box formulation in the local frame
includes  tidal gravity and Coriolis force terms in the horizontal direction.
In addition, we include a (fixed) vertical gravitational potential to model the
stellar disk and dark matter halo, self-gravity of gas, cooling and heating,
and thermal conduction.  The resulting set of equations (see e.g.,
\citealt{pio07}; Paper~I) is
\begin{equation}\label{eq:cont}
\pderiv{\rho}{t}+\divergence{\rho\vel}=0,
\end{equation}
\begin{equation}\label{eq:mom}
\pderiv{\vel}{t}+\advect{\vel}= -\frac{1}{\rho}\nabla{P}-2\Om\times\vel
+2q\Omega^2 x \xhat -\nabla\Phi+\mathbf{g}_{\rm sd},
\end{equation}
\begin{equation}\label{eq:energy}
\pderiv{e}{t}+\divergence{e\vel}=-P\nabla\cdot{\vel}
-\rho\mathcal{L}+\mathcal{K}\nabla^2 T,
\end{equation}
\begin{equation}\label{eq:poisson}
\nabla^2\Phi=4\pi G\rho,
\end{equation}
where $\mathbf{g}_{\rm sd}$ is the external gravity, $\rho\mathcal{L}$ is the
net cooling rate per unit volume, and
$\mathcal{K}=4\times10^7\ergs\cm^{-1}\Kel^{-1}/[1+(0.05\pcc/n)]$ is the
conductivity adopted such that thermal instability is resolved on our grid
(\citealt{koy09a}, Paper I).  Other symbols have their usual meanings.  We
assume that the gas has cosmic abundance so that the gas pressure is
$P=1.1n\kbol T$ where $n=\rho/(1.4 m_p)$ is the number density of hydrogen
nuclei. An ideal gas law is assumed with internal energy density $e=(3/2)P$.

Since the vertical gradient scales of stellar disks and dark matter halos are
generally much larger than those of gaseous disks (and vertical domain sizes in
our simulations), we adopt the simple approach of taking external gravity as a
linear function of the vertical coordinate $z$:
\begin{equation}\label{eq:gsd}
\mathbf{g}_{\rm sd}=-4\pi G\rhosd z \zhat,
\end{equation}
where $\rhosd$ is the stellar {\it plus} dark matter volume density at the
midplane.

The net volumetric cooling function is given by $\rho\mathcal{L}\equiv
n[n\Lambda(T)-\Gamma]$.  For the diffuse ISM, radiative cooling by the
\ion{C}{2} 158$\mu$m fine-structure line and by Ly$\alpha$ line emission is
dominant at low and high temperature, respectively, whereas grain photoelectric
heating by FUV radiation with energy $6\eV<h\nu<13.6\eV$ is dominant in both
the cold and warm phases \citep{bak94}.  We adopt the fitting formula for the
cooling function from \citet{koy02}:
\begin{eqnarray}\label{eq:cool}
\Lambda(T)&=&2\times 10^{-19}\exp\left(\frac{-1.184\times10^5}{T+1000}\right)\nonumber\\
&& +2.8\times10^{-28}\sqrt{T}\exp\left(\frac{-92}{T}\right)
{\rm \;erg} \cm^3 \;{\rm s^{-1}},
\end{eqnarray}
with temperature $T$ in degrees Kelvin.

We adopt the same star formation feedback prescription as in Paper~I.  This
includes momentum feedback (to represent the radiative stage of blasts produced
by SN explosions) at a rate proportional to the SFR, together with a
time-dependent heating rate that is also proportional to the SFR (representing
radiative heating from young, massive stars).  We scale the heating rate
relative to the fiducial Solar-neighborhood value adopted by \citet{koy02},
$\Gamma_0=2\times10^{-26}\ergs$ (see below).

The reader is referred to Paper~I for a full description of the feedback
prescription. Here we only give brief summary. Star formation is taken to occur
only when the density of grid zone exceeds a threshold\footnote{ The threshold
density is calculated based on the  thermal equilibrium state of cold gas,
$\nth=\Gamma/\Lambda(\Tth)$, where $\Tth$ is the threshold temperature.  For
our net cooling function, the Jeans length is $\lambda_J \approx
1.4\Tth^{3/4}e^{-46/\Tth} (\Gamma/\Gamma_0)^{-1/2} \pc$ (Paper~I).  By taking
$\lambda_J=5\pc$, $\Tth$ (and hence $\nth$) can be found as a function of
$\Gamma/\Gamma_0$.  Using a power-law fit we obtain $\nth$ given in the main
text, representing the maximum density for which the Jeans length is well
resolved \citep[cf.][]{1997ApJ...489L.179T} at our adopted spatial resolution
of $2\pc$.} $\nth=200\pcc(\Gamma/\Gamma_0)^{0.2}$.  The probability of massive
star formation in a given zone with simulation time step $\Delta t$ is
\begin{equation}\label{eq:nsn}
\mathcal{P}_{\rm *}=\frac{\dot{M}_*}{\Msn}\Delta t,
\end{equation}
where $\dot{M}_*$ is the expected SFR for a given grid zone, and
$\Msn=100\Msun$ represents the total mass in stars (averaged over the IMF) per
SN \citep[cf.,][]{kro01}.  The expected SFR for a grid zone with density $\rho
> \rhoth$ is taken as
\begin{equation}\label{eq:sfr}
\dot{M}_*=\dot{\rho}_*\Delta V=\eff\frac{\rho\Delta V}{\tff(\rho)},
\end{equation}
where $\Delta V= \Delta x \Delta y \Delta z =(2\pc)^3$ is the volume element of
the grid zone, $\eff$ is the star formation efficiency per free-fall time of
dense, self-gravitating gas, and $\tff(\rho)\equiv[3\pi/(32G\rho)]^{1/2}$ is
the free-fall time of a given grid zone. We take a fiducial value $\eff=0.01$
consistent with theory and observations of dense ISM gas
\citep{kru05,kru07,kru12}.  We note that based on the study of SO12, the
adopted value of $\eff$ can be varied by an order of magnitude with little
effect on the resulting mean SFR, provided that the threshold density is
sufficiently high compared to the mean value in the diffuse ISM.

When a massive star forms, we immediately apply momentum feedback in its
vicinity, i.e. we neglect time delays (see below and Paper~I).   For each
feedback event, we first take spatial averages within a sphere of radius $\rsh$
and redistribute mass, momentum, and thermal energy with their respective
averaged values.  We adopt a fixed value $\rsh=10\pc$ as representative of the
shell formation epoch \citep{cio88,koo04}. We then add to the local momentum
density a quantity with a form $\rho \mathbf{v}(\mathbf{r}) = p_{\rm sh}
(r/\rsh)^2\hat{\mathbf{r}}$, where $\mathbf{r}$ is the position vector measured
from the center of the SN feedback region, and $p_{\rm sh}=5\Psn/(4\pi\rsh^3)$
is the momentum density at $r=\rsh$.  This injects a total radial momentum
$\Psn$ to the surrounding medium; physically, $\Psn$ represents the value of
the shell momentum in the  radiative stage of a supernova remnant.
\citet{tho98} studied evolution of expanding spherical supernova remnants with
realistic radiative cooling and found that the  shell momentum is 
$\Psn\sim(1-4) \times 10^5 \Msun\kms$ at the time of maximum luminosity and a
factor $\sim 2.5$ larger after shell cooling has declined,  with the highest
values corresponding to low metallicity.  The simulations of \citet{cio88} and
\citet{1998ApJ...500..342B} found similar values at Solar metallicity.  In the
present work, we adopt $\Psn=3\times10^5\Msun\kms$.  For uniform-density
conditions, there is only a weak dependence of the radiative-stage momentum on
the ambient density ($\Psn \propto \bar n^{-0.12}$), because momentum in the
Sedov stage varies as $\sim E_{\rm SN}/v_{\rm shock}$ and the onset of the
radiative stage is the point at which $v_{\rm shock}$ drops enough that 
post-shock cooling becomes strong  (see e.g. the physical discussion in
\citealt{1998ApJ...500..342B}).  For a strongly clumped medium, however, both
the radius at the radiative stage and the net momentum injection may depart
more strongly from our adopted value; this will be evaluated with future
simulations.

For radiative feedback, we count the number of recent massive star formation
events over $t_{\rm bin}$ to calculate the recent SFR:
\begin{equation}\label{eq:sfrsurf}
\SigSFR=\frac{\mathcal{N}_{*}\Msn}{L_x L_y t_{\rm bin}},
\end{equation}
where $\mathcal{N}_{*}$ stands for the total number of massive stars formed
during the time interval ($t-t_{\rm bin}$, $t$).  In contrast to Paper~I, our
simulation domain is sufficiently large in the azimuthal direction so that
$t_{\rm bin}$ can be set to the realistic lifetime of OB stars, $t_{\rm
FUV}=10\Myr$ \citep{par03}.\footnote{ For the QA02 model, which has extremely
low surface density, we adopt $t_{\rm bin}=40\Myr$ to partly compensate for the
fact that the domain of influence of FUV radiation would be approximately four
times larger than the horizontal area of our simulation box.  Note that this
time is still very short compared to the orbit time $\torb$ and the vertical
oscillation period $t_{\rm osc}$ for the QA02 model.  Also, for 3DS and XZ
models with smaller azimuthal domain size (see Section \ref{sec:model}), we
extend $t_{\rm bin}$ to $0.5\torb$.} We assume a simple linear relationship
between the heating rate $\Gamma$, the mean FUV radiation field $J_{\rm FUV}$,
and the SFR surface density $\SigSFR$ (see OML10; Paper~I) normalized relative
to Solar neighborhood conditions, such that
\begin{equation}\label{eq:heat}
\Gamma=\Gamma_0\sbrackets{\frad\rbrackets{\frac{\SigSFR}{\SigSFRsun}}
+\rbrackets{\frac{J_{\rm FUV,meta}}{J_{\rm FUV,0}}}}.
\end{equation}
In Equation (\ref{eq:heat}), we adopt Solar neighborhood fiducial heating rate
$\Gamma_0=2\times10^{-26}\ergs$ from \citet{koy02}, SFR surface density in the
Solar neighborhood $\SigSFRsun=2.5\times10^{-3}\sfrunit$ from \citet{fuc09},
and $J_{\rm FUV,meta}=0.0024 J_{\rm FUV,0}$ to represent the metagalactic FUV
radiation field \citep{ste02}.  As in Paper~I, we introduce a parameter $\frad$
to allow for variable heating efficiency at a given SFR relative to our adopted
parameters.  By increasing/decreasing $\frad$, we can also represent
greater/lesser penetration of FUV through the ISM as would occur for
lower/higher dust abundance.

We utilize the Athena code with the van Leer integrator \citep{sto09}, Roe's
Riemann solver, a piecewise linear spatial reconstruction scheme, and the
orbital advection method for a shearing box \citep{sto10}. In addition, we
adopt an FFT Poisson solver method with treatment of shearing horizontal
coordinates as introduced by \citet{gam01} and with vacuum vertical boundary
conditions as introduced by \citet{koy09a}.  We include cooling/heating and
thermal conduction terms in an operator split manner.  For the net cooling, we
use an implicit solution method based on Simpson's rule and apply subcycling to
limit the maximum temperature change to $<50\%$ of previous value over all grid
zones. If the change of temperature exceeds $50\%$ of the previous value, we
halve the timestep and repeat subcycles for the specific grid zone until the
temperature change for one hydrodynamic timestep update is smaller than $50\%$.

Similar to Paper~I, we mention several caveats regarding the current
simulations. First, the feedback from SN explosions is realized solely via
expanding SN remnants rather than injecting thermal energy, which would create
a hot ISM \citep{mck77}. More realistically, the hot ISM would occupy a
significant volume even near the midplane \citep[e.g.][]{hil12}, together with
the cold and warm phases in approximate pressure equilibrium.  Due to its large
scale height, however,  most of the hot gas does not participate in supporting
the weight of the warm/cold diffuse gas, although expansion of highly
overpressured hot SN remnants   is crucial in driving turbulence within the
surrounding  warm/cold ISM; we model the latter effect.  Based on our
conclusion that the SFR is regulated by midplane total pressure (Section
\ref{sec:scaling}), we believe that our main findings are robust, in spite of
our simplified treatment of SNe.  We note that using preliminary simulations in
which SNe are modeled with thermal energy input (rather than momentum input),
we recover similar results for SFRs to those reported here.

A second caveat is that the feedback we apply is instantaneous, whereas a more
realistic treatment of stellar evolution would include inputs from stellar
winds and expanding \ion{H}{2} regions \citep[e.g.,][]{mac04} prior to SN
explosions.  Since SNe are the most powerful driving source of turbulence
averaged over the ISM ($\Psn/\Msn$ for stellar winds and expanding \ion{H}{2}
regions would be an order of magnitude lower than that of SNe; see OS11), we
believe that the current simplified approach is an adequate first approximation
for modeling SFR self-regulation in diffuse-dominated regions.  Feedback from
earlier stages of massive star evolution would, however, affect the detailed
properties and lifetimes of gravitationally-bound clouds (GBCs), so including
these effects will be important in modeling higher-$\Sigma$ galactic
regions where most of the ISM mass is in GBCs rather than diffuse structures.
For this reason, we confine ourselves here to the regime $\Sigma \le 20 \Surf$
in which the observed ISM is predominantly in the diffuse atomic component.
When spiral arms or bars are taken into account, the time delay between the
epochs of star formation and feedback would likely also be important
\citep[e.g.,][]{kko10,seo13}.

\begin{deluxetable}{lcccccc}
\tabletypesize{\footnotesize} \tablewidth{0pt}
\tablecaption{Model Parameters\label{tbl:model}}
\tablehead{
\colhead{Model} &
\colhead{$\Sigma$} &
\colhead{$\rho_{\rm sd}$} &
\colhead{$\torb$} &
\colhead{$L_z$} &
\colhead{$s_0$} &
\colhead{$f_{\rm rad}$} \\
\colhead{}&
\colhead{[$\Msun\,\pc^{-2}$]}&
\colhead{[$\Msun\,\pc^{-3}$]}&
\colhead{[$\Myr$]}&
\colhead{[$\pc$]} &
\colhead{} &
\colhead{}
}
\startdata
\bf{QA02} &   2.5 &   0.0031 &   878 &  2048 & 0.28 &  1.0 \\
\bf{QA05} &   5.0 &   0.0125 &   439 &  1024 & 0.28 &  1.0 \\
QA07 &   7.5 &   0.0281 &   293 &   768 & 0.28 &  1.0 \\
\bf{QA10} &  10.0 &   0.0500 &   219 &   512 & 0.28 &  1.0 \\
QA15 &  15.0 &   0.1125 &   146 &   384 & 0.28 &  1.0 \\
\bf{QA20} &  20.0 &   0.2000 &   110 &   256 & 0.28 &  1.0 \\
\hline
QB02 &   2.5 &   0.0125 &   878 &  1024 & 0.07 &  1.0 \\
QB05 &   5.0 &   0.0500 &   439 &   768 & 0.07 &  1.0 \\
QB07 &   7.5 &   0.1125 &   293 &   512 & 0.07 &  1.0 \\
QB10 &  10.0 &   0.2000 &   219 &   384 & 0.07 &  1.0 \\
QB15 &  15.0 &   0.4500 &   146 &   256 & 0.07 &  1.0 \\
\hline
S02  &   2.5 &   0.0500 &   878 &  1024 & 0.02 &  1.0 \\
S07  &   7.5 &   0.0500 &   293 &   768 & 0.16 &  1.0 \\
S15  &  15.0 &   0.0500 &   146 &   512 & 0.62 &  1.0 \\
S20  &  20.0 &   0.0500 &   110 &   512 & 1.10 &  1.0 \\
\hline
G01  &  10.0 &   0.0125 &   219 &  1024 & 1.10 &  1.0 \\
G02  &  10.0 &   0.0250 &   219 &   768 & 0.55 &  1.0 \\
G10  &  10.0 &   0.1000 &   219 &   512 & 0.14 &  1.0 \\
G40  &  10.0 &   0.4000 &   219 &   384 & 0.03 &  1.0 \\
\hline
R02  &  10.0 &   0.0500 &    28 &   512 & 0.28 &  0.2 \\
R05  &  10.0 &   0.0500 &    28 &   512 & 0.28 &  0.5 \\
R25  &  10.0 &   0.0500 &    28 &   512 & 0.28 &  2.5 \\
\bf{R50}  &  10.0 &   0.0500 &    28 &   512 & 0.28 &  5.0
\enddata
\tablecomments{ Physical input parameters are the same as in  Paper~I. Full 3D
simulations with $L_x=L_y=512\pc$ are run only for models QA02, QA05, QA10,
QA20, and R50 (bold face in first column). ``Slim'' 3D simulations
($L_x=512\pc$, $L_y=32\pc$) are run for all parameters.}
\end{deluxetable}

\subsection{Model Parameters}\label{sec:model}
We run the same set of models as in Paper~I, which covers a wide range of
outer disk conditions for nearby galaxies.  Our parameters are: the gas surface
density $\Sigma$, the stellar plus dark matter density at the midplane
$\rhosd$, and the galactic rotational speed $\Omega$.  We have five model
series: QA, QB, S, G, and R.  For all series, the angular speed of galactic
rotation varies as $\Omega=28\kms\kpc^{-1}(\Sigma/10\Msun\pc^{-2})$ such that
the gaseous Toomre stability parameter $Q_g\equiv\kappa\sigma_x/(\pi G\Sigma)$
would be constant for fixed radial ($\hat x$) gas velocity dispersion
$\sigma_x$ ($Q_g\sim2$ for $\sigma_x=7\kms$).  In the QA and QB series,
$\rhosd\propto\Sigma^2$ such that the stellar Toomre parameter would also be
constant within each series.  These two series differ only in the ratio of
self-to-external (i.e.  gaseous-to-stellar+dark matter) gravity: $s_0=0.28$ and
$0.07$ for the QA- and QB-series, respectively, where $s_0\equiv\pi
G\Sigma^2/(2\sigma_z^2\rhosd)$ \citep[cf.,][]{kim02a}. The QA and QB series may
thus each be thought of as representing a sequence of radii in a $Q_g=const$,
$Q_*=const$ galaxy, where the stellar disk is a factor of four more massive in
the QB series than in the QA series.  For the S series we fix $\rhosd$ and vary
$\Sigma$, whereas for the G series we fix $\Sigma$ and vary $\rhosd$. For the R
series, we vary $\frad$ for the fiducial model QA10, to test the effect of
varying the heating efficiency or dust shielding for FUV.  We list the model
parameters in Table~\ref{tbl:model}.  In all models, the orbital period is
$\torb=2\pi/\Omega=220\Myr (\Omega/28\kms\kpc^{-1})^{-1}=220\Myr
(\Sigma/10\Surf)^{-1}$, which we take as the time unit in our presentation.

We take $L_x=512\pc$ and $L_z=4H_w$ for the horizontal and vertical domain
sizes. Here $H_w\equiv c_w/(4\pi G\rhosd)^{1/2}$ is a nominal Gaussian scale
height of warm gas with $c_w=7\kms$, which varies from model to model.  In all
models, we vary the number of grid zones such that the grid
resolution\footnote{Although the resolution is a factor of two lower than in
Paper~I, we have confirmed that the key physical properties are converged even
at lower resolution than we adopt here.} is $\Delta x=\Delta y = \Delta z =
2\pc$.  In the azimuthal direction, we consider two different domain sizes: one
set is full 3D simulations using azimuthal domain size $L_y=L_x=512\pc$
(hereafter 3DF models), and the other set uses a slimmer azimuthal domain size
$L_y=32\pc$ (hereafter 3DS models).  Since full 3D simulations require
considerable computational resources, it is impractical to run 3DF models for
all parameter values.  Our 3DS models cover the whole parameter space, while
the 3DF models cover just the QA and R series (see Table~\ref{tbl:model}).  In
forthcoming sections, we shall show that 3DF and 3DS models yield essentially
the same results in terms of statistical properties at saturation.  We also
compare these properties to the results from the simulations of Paper~I, which
followed evolution of two-dimensional radial-vertical slices through the disk;
these are denoted as ``XZ'' models.  In the remainder of this paper, we use
suffixes 3DF, 3DS, and XZ to distinguish models with the same parameters but
different azimuthal domain size.  The term ``3D models'' denotes both 3DF and
3DS models.

\section{Simulation Results}\label{sec:result}
\subsection{Overview of Time Evolution and Disk Properties}\label{sec:evol}

\begin{figure*}
\epsscale{1.0}
\plotone{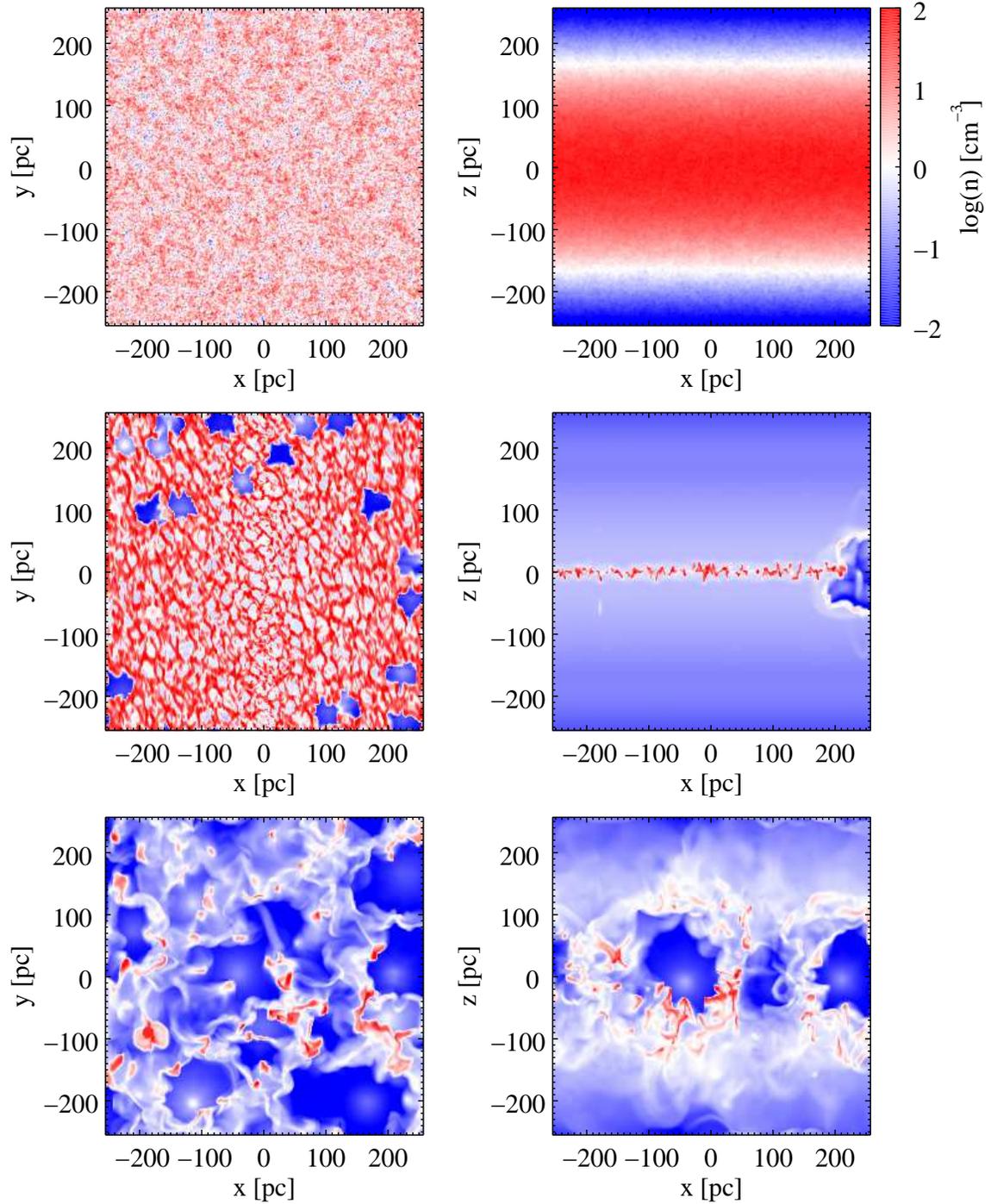}
\caption{Snapshots of density in the Solar-neighborhood-like model QA10-3DF
(logarithmic color scale) at early evolutionary stages, $t/\torb=0$ (top),
$0.1$ (middle), and $0.2$ (bottom).  Left and right columns display horizontal
and vertical slices through the computational domain at $z=0$ and $y=0$,
respectively. The initial disk rapidly separates into two phases due to thermal
instability, with the cold gas settling to the midplane \citep{kko10}.  Gravity
causes the cold gas to collect into GBCs where star formation occurs, and
produces feedback, starting at $t/\torb=0.1$ (middle row).  Energy injected by
SNe drives turbulence, expanding the disk vertically (see evolution from middle
right to lower right) and helping to create large-scale clumpy/filamentary
structure.  \label{fig:QA10_early}}
\end{figure*}

\begin{figure}
\epsscale{0.9}
\plotone{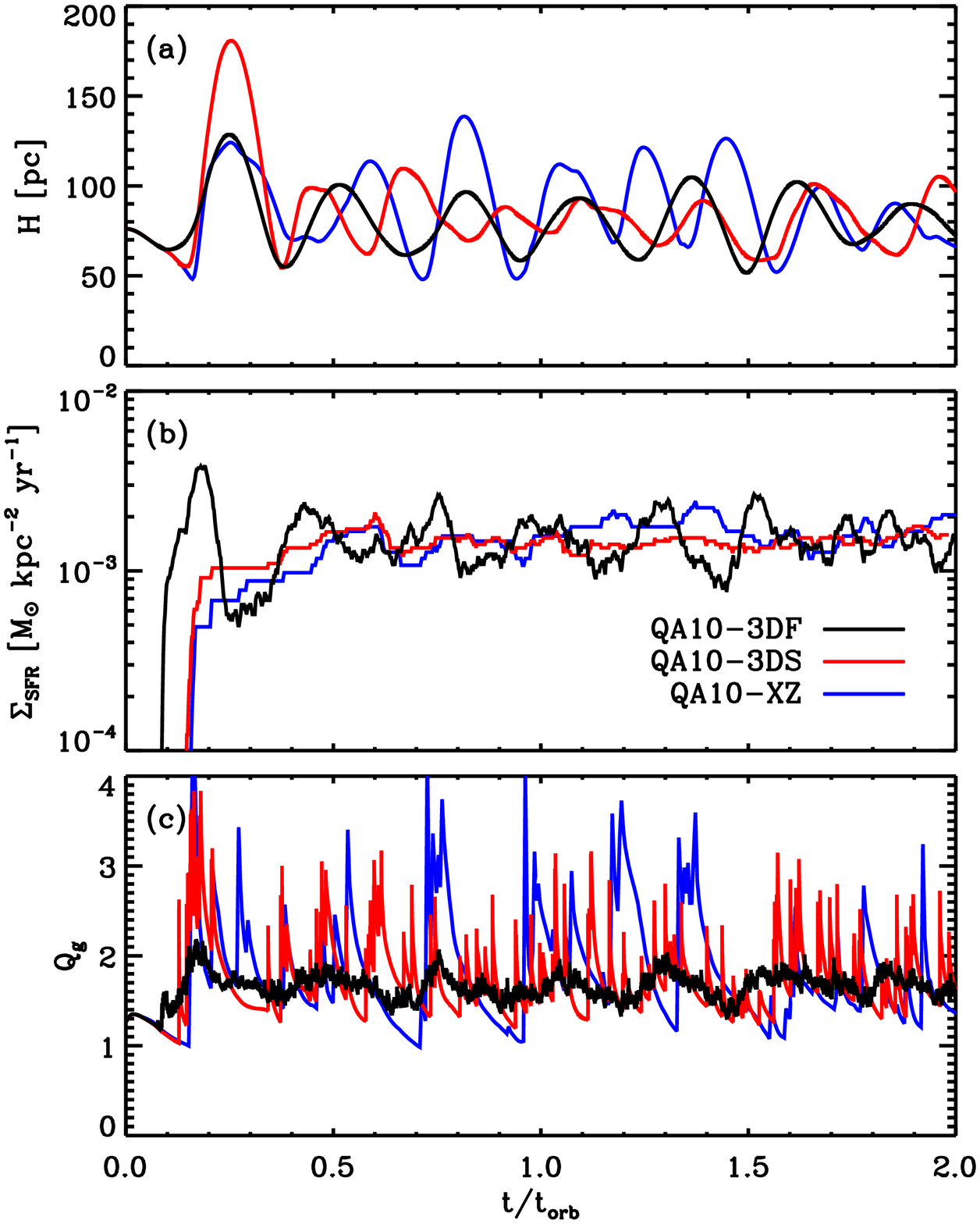} \caption{ Time evolution of
(a) the density-weighted vertical scale height $H$, (b) the SFR surface density
$\SigSFR$, and (c) the gaseous Toomre stability parameter $Q_g$ for models
QA10-3DF (black), QA10-3DS (red), and QA10-XZ (blue). These models differ only
in their azimuthal domain sizes.  The evolution is similar for all models. The
temporal fluctuations in $\SigSFR$ are smeared out in models QA10-3DS and
QA10-XZ due to the larger binning adopted for $\SigSFR$ (see text).  The
temporal fluctuations in $Q_g$ are smeared out in model QA10-3DF due to a much
larger averaging volume compared to models QA10-3DS and QA10-XZ.  The mean
values of $H\sim80\pc$, $\SigSFR\sim1.5\times10^{-3}\sfrunit$, and $Q_g\sim1.7$
after saturation ($t/\torb>1$) are essentially the same in all models (see
Table \ref{tbl:stat1}).  \label{fig:hst1}}
\end{figure}

In this subsection, we describe details of time evolution and properties of our
disk models.  We begin with our fiducial model, QA10, which adopts
$\Sigma=10\Surf$, $\rhosd=0.05\rhounit$, and $\Omega=28\kms\kpc^{-1}$, similar
to conditions in the Solar neighborhood.  Figure~\ref{fig:QA10_early} shows
evolving density slices of the QA10-3DF model in the horizontal XY-plane (at
the disk midplane $z=0$; left) and radial-vertical XZ-plane ($y=0$; right) at
$t/\torb=0$ (top), $0.1$ (middle), and $0.2$ (bottom).  The initial gas
distribution (top row of Figure~\ref{fig:QA10_early}) follows a Gaussian
vertical density profile with scale height of $\sim80\pc$. In the initial
conditions, we impose a Gaussian random density perturbation with flat spectrum
at wavenumbers smaller than $kL_z/2\pi=8$ and total amplitude of $10\%$. This
choice of initial conditions allows rapid growth of thermal instability near
the midplane, evolving towards two-phase thermal equilibrium
\citep[see][]{kko10}.\footnote{Although the specific initial conditions affect
the initial model evolution, evolution at later stages and the resulting
saturated-state statistical properties are similar irrespective of the choice
of initial conditions.} At the midplane, thermal instability forms cold
cloudlets, and subsequent gravitational accretion leads to growth of more
massive clouds.  The first star formation and SN feedback event is triggered at
$t\sim0.1 \torb\sim 20 \Myr$ (see middle row of Figure~\ref{fig:QA10_early}),
and many subsequent events follow.  The feedback events disperse cold cloudlets
and swell the gas disk vertically (see bottom row of
Figure~\ref{fig:QA10_early}).  Dispersed gas slows as it climbs vertically in
the combined potential of stars and gas, and then falls back to the midplane.
Gravitational condensation of clouds into larger structures leads to new
high-density regions with subsequent star formation and feedback events.
Driven by these processes, the gas disk undergoes a quasi-periodic cycle of
vertical ``breathing'' oscillations with period of $t_{\rm osc}\sim
0.5(\pi/G\rhosd)^{1/2}$ (see Figure~\ref{fig:hst1}), equal to $\sim 60$Myr for
model QA10.  This is half of the free-particle vertical oscillation period
because cloudlets collide at the midplane.

Figure~\ref{fig:hst1} presents the time histories of (a) the disk scale height
$H\equiv [\int\rho z^2 dV/\int\rho dV]^{1/2}$, (b) the SFR surface density
$\SigSFR$, and (c) the gaseous Toomre stability parameter $Q_g \equiv \kappa
\sigma_x/\pi G \Sigma $, with $\sigma_x^2 \equiv \int (P + \rho v_x^2) dV /\int
\rho dV$, for models QA10-3DF (black), QA10-3DS (red), and QA10-XZ (blue).
All quantities reach quasi-steady saturated values after one orbit, implying
that statistical measures can be computed starting at this epoch.  All three
models, with different azimuthal domains, are overall in very good agreement
with each other, confirming the reliability of our previous XZ models and the
3DS models for the purposes of assessing mean values of $H$ and $\SigSFR$.  The
scale height, the SFR surface density, and the Toomre parameter have mean
values $\abrackets{H}=81$, $80$, and $88\pc$, $\abrackets{\SigSFR}=1.5$, $1.5$,
and $1.7\times10^{-3}\sfrunit$, and $\abrackets{Q_g}=1.7$, $1.8$, and $1.8$
for QA10-3DF, QA10-3DS, and QA10-XZ models, respectively. The angle brackets
$\abrackets{}$ denote a time average over $t/\torb=1-2$ (note that in Paper~I,
time averages are taken for $t/\torb=2-3$).  Since QA10-3DS and QA10-XZ models
adopt $\tbin=0.5\torb>t_{\rm osc}$ for the purpose of computing $\SigSFR$ and
heating rates,  the temporal fluctuation in $\SigSFR$ is reduced for these
models.  Similarly, the spiky profiles shown in QA10-3DS and QA10-XZ models are
not seen in QA10-3DF model due to the larger spatial averaging volume.

\begin{figure}
\epsscale{1.0}
\plotone{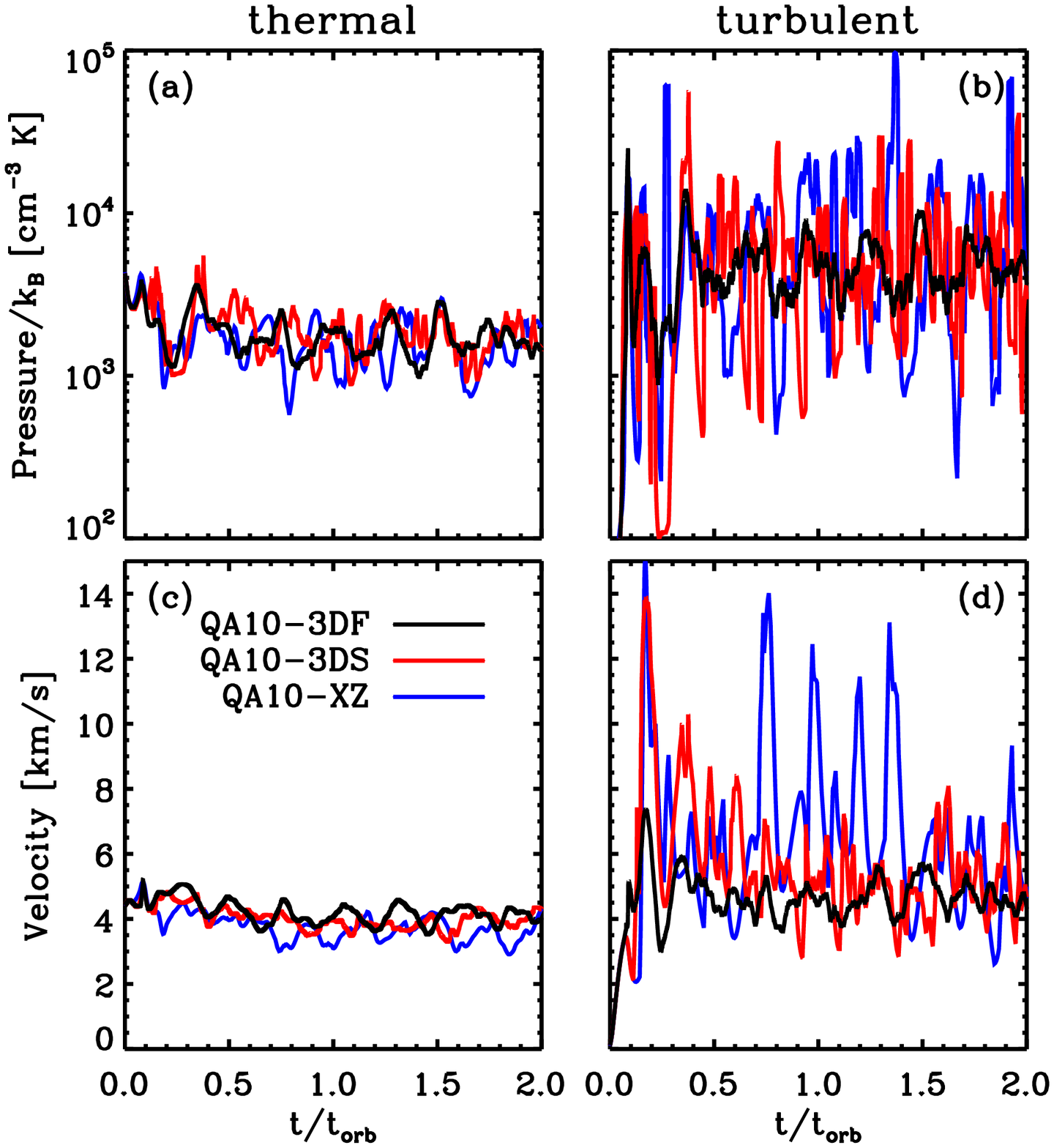}
\caption{ Time evolution of (a) thermal and (b) turbulent pressures at the
midplane, together with the mass-weighted (c) thermal and (d) turbulent
velocity dispersions of the diffuse component for models QA10-3DF (black),
QA10-3DS (red), and QA10-XZ (blue).  Boxcar averages with a window of $\Delta
t=0.02\torb$ are taken to reduce noisy spikes and show fluctuations clearly.
\label{fig:hst2}}
\end{figure}

\begin{figure}
\epsscale{1.0}
\plotone{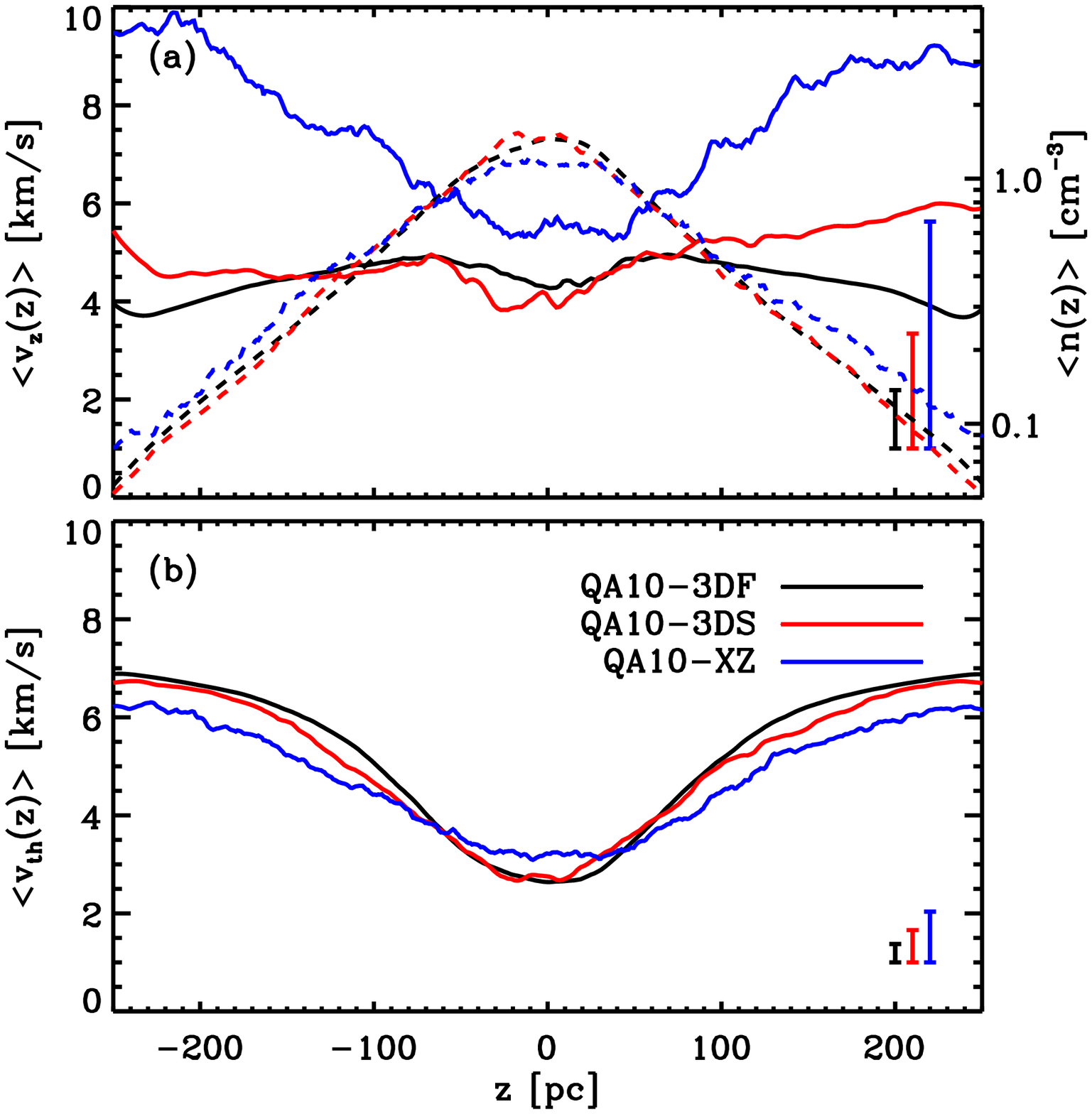}
\caption{ Temporally and horizontally averaged velocity dispersion profiles
(solid curves) in $z$ (see equation~(\ref{eq:vzprof})) for (a) vertical
turbulent and (b) thermal components.  The mean standard deviations of temporal
fluctuations are indicated as errorbars in the bottom-right corner.  In (a),
the dashed curves and right axis show the corresponding density profiles.
\label{fig:zprof}}
\end{figure}

Figure~\ref{fig:hst2} plots the time histories at the midplane of (a) thermal
and (b) turbulent pressure in all QA10 models.  As demonstrated in Paper~I, the
total pressure at the midplane, $\Ptot$, is crucial because it must match the
vertical weight of the ISM (i.e. dynamical equilibrium pressure), and because
it is also directly related to $\SigSFR$ via feedback.  Since $\Ptot$ consists
of both thermal and turbulent components, which are independently connected to
the SFR (see Section~\ref{sec:equil} and Paper~I), it is useful to calculate
the thermal and turbulent pressures separately by taking volume-weighted
horizontal averages as
\begin{equation}\label{eq:Pth}
\Pth = \frac{\int_{z=-\Delta z/2}^{z=+\Delta z/2}\int\int P\Theta(n\!<\!\ngbc) dxdy
dz} {\int_{z=-\Delta z/2}^{z=+\Delta z/2}\int\int\Theta(n\!<\!\ngbc)dxdy dz},
\end{equation}
\begin{equation}\label{eq:Pturb}
\Pturb = \frac{\int_{z=-\Delta z/2}^{z=+\Delta z/2}\int\int \rho
v_z^2\Theta(n\!<\!\ngbc) dxdy dz} {\int_{z=-\Delta z/2}^{z=+\Delta
z/2}\int\int\Theta(n\!<\!\ngbc)dxdy dz}.
\end{equation}
Here, to average only over diffuse gas, $\Theta(X)$ is 1 if the logical
argument `X' is true and 0 otherwise.  We choose $\ngbc\equiv 50\pcc$ as the
minimum density for GBCs which are compressed by self-gravity to higher thermal
pressure than their surroundings.\footnote{We have checked that the exact
choice of $\ngbc$ does not significantly affect our statistical results for
diffuse gas provided it is high enough to safely separate out high-pressure
cold gas.} In Figure~\ref{fig:hst2}, we have applied a boxcar average with
window size of $\Delta t=0.02\torb$ to show time evolution clearly.  The mean
thermal and turbulent pressures at the midplane are quite similar among all
QA10 models (see also Columns (3) and (4) of Table~\ref{tbl:stat1}).  Turbulent
pressures in the QA10-3DS and QA10-XZ models fluctuate with larger amplitude
than in the QA10-3DF model because of their smaller averaging domains.

Figure~\ref{fig:hst2}(c) and (d) respectively show the mass-weighted thermal
and turbulent vertical velocity dispersions of the diffuse component, given by
\begin{eqnarray}\label{eq:veld}
\vthdiff\equiv\rbrackets{\frac{\int P\Theta(n\!<\!\ngbc) dxdydz}
{\int\rho\Theta(n\!<\!\ngbc) dxdydz}}^{1/2}, \nonumber \\
\vzdiff\equiv\rbrackets{\frac{\int \rho v_z^2\Theta(n\!<\!\ngbc) dxdydz}
{\int\rho\Theta(n\!<\!\ngbc) dxdydz}}^{1/2}.
\end{eqnarray}
Since the mass-weighted velocity dispersions averaged over the whole simulation
volume relate most closely to quantities that can be directly observed, we use
these velocities as observational proxies in our simulations.  The
mass-weighted thermal velocity dispersions are $\sim4\kms$ for all QA10 models,
while the turbulent velocity is higher in model QA10-XZ
($\abrackets{\vzdiff}=7.2\pm2.3\kms$) than in models QA10-3DS
($\abrackets{\vzdiff}=5.1\pm1.0\kms$) and QA10-3DF
($\abrackets{\vzdiff}=4.7\pm0.5\kms$).

Since the midplane values of both thermal and turbulent pressures are in good
agreement among all QA10 models, the differences of the mass-weighted turbulent
velocity dispersions, which are averaged over the whole simulation domain, must
arise from differences in the vertical profiles.  To see this clearly, we
define the mass-weighted, horizontally-averaged velocity dispersions as
functions of $z$:
\begin{eqnarray}\label{eq:vzprof}
v_{\rm th}(z)={\sbrackets{\frac{\int P\Theta(n\!<\!\ngbc)
dxdy}{\int\rho\Theta(n\!<\!\ngbc) dxdy}}^{1/2}}, \nonumber\\
v_z(z)={\sbrackets{\frac{\int \rho
v_z^2\Theta(n\!<\!\ngbc)dxdy}{\int\rho\Theta(n\!<\!\ngbc) dxdy}}^{1/2}}.
\end{eqnarray}
Figure~\ref{fig:zprof} plots vertical profiles of (a) $\abrackets{v_{\rm
th}(z)}$ and (b) $\abrackets{v_{z}(z)}$ based on time-averages for QA10-3DF
(black), QA10-3DS (red), and QA10-XZ (blue) models.  The vertical profiles of
thermal velocity dispersion $\abrackets{v_{\rm th}(z)}$ show similar trends for
all QA10 models, increasing as $|z|$ increases since the warm gas dominates at
high-$|z|$.  However, the vertical profiles of turbulent velocity dispersion
$\abrackets{v_z(z)}$ in QA10-3D models are nearly flat or even decrease at high
$|z|$, while $\abrackets{v_z(z)}$ in the QA10-XZ model secularly increases with
$|z|$.  This is presumably because the total mass swept up by an expanding
shell is larger in 3D models (spherical volume $\propto r^3$) than in XZ models
(cylindrical volume $\propto r^2 \rsh$ for $L_y=2\rsh$).  Although the feedback
is normalized for the XZ models such that the injected momentum is the same
as for 3D models (see Paper~I), the resulting turbulent velocities at high
$|z|$ are smaller in 3D because the larger swept-up mass in 3D reduces the mean
velocity at large $|z|$.  Near the midplane, at $|z|\simlt H$ (i.e. where
density is within a factor $\sim 3$ of the midplane value; see right axis in
Fig. \ref{fig:zprof}(a)), turbulent velocity dispersions for XZ models are
similar to those for 3D models.

\begin{figure}
\epsscale{1.0}
\plotone{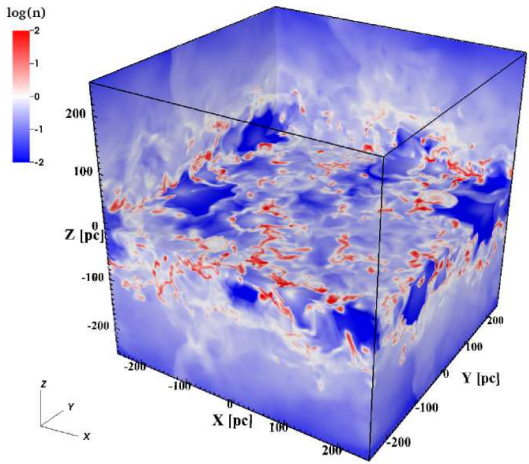}
\caption{ Density structure in Solar neighborhood model  QA10-3DF at
$t/\torb=1.12$, after a fully turbulent state is reached.  Colorbar labels
$\log n$ (cm$^{-3}$) in five different planes ($x=\pm L_x/2$, $y=\pm L_y/2$,
and $z=0$).  Cloudy/filamentary structure is evident, as well as dispersal of
dense gas above and below the midplane by feedback-driven turbulence.
\label{fig:QA10_3D}}
\end{figure}

\begin{figure*}
\epsscale{1.0}
\plotone{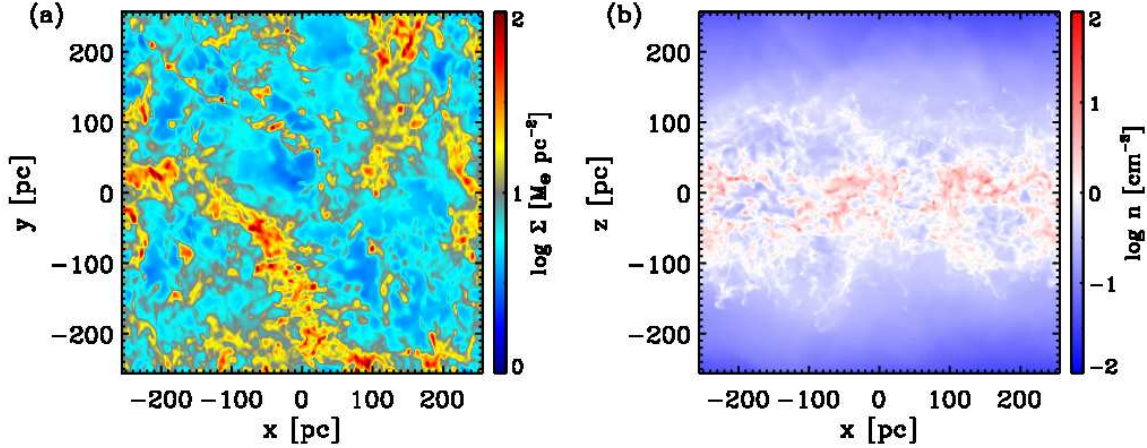}
\caption{ (a) Surface density projected on the horizontal plane for the same
snapshot shown in Figure \ref{fig:QA10_3D}.  The radial direction is along $x$,
and the azimuthal direction is along $y$.  Colorbar labels $\log \Sigma$
($\Surf$).  (b) Radial-vertical density structure based on azimuthal (i.e.
along $y$) average of the same snapshot shown in Figure \ref{fig:QA10_3D}.
Colorbar labels $\log(n)$ ($\cm^{-3}$).  \label{fig:QA10proj}}
\end{figure*}

Figure~\ref{fig:QA10_3D} shows slices through the volume for a snapshot
($t/\torb=1.12$) of model QA10-3DF after a quasi-steady state is reached.  The
structure of the ISM is filamentary, with high porosity and relatively
large-scale structures due to the combined action of SN events and
self-gravity.  Some high density cloudlets are located near SN shell
boundaries, as shown in Paper~I, although cloudlets are also present far from
these shells.  In Figure~\ref{fig:QA10proj}, we show (a) the column density of
gas projected onto the horizontal plane and (b) the volume density averaged
along the azimuthal ($y$) direction, for the same snapshot.  Although
Figure~\ref{fig:QA10proj}(a) contains a large-scale diagonal feature,
examination of model animations over several orbits shows that this kind of
sheared structure grows and then disperses before reaching very large
amplitude.  This is likely because continuous kinetic energy input from SN
feedback keeps the turbulent velocity dispersion large enough to maintain $Q_g$
within the range between 1.5 and 2 (see Figure~\ref{fig:hst1}(c)) in which
swing amplification is not strong \citep[e.g.,][]{kim01,kim02a,kim07}.

\begin{figure}
\epsscale{1.0}
\plotone{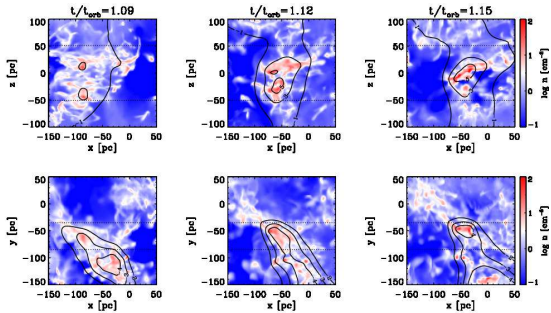}
\caption{ Example of cloud growth by gravitationally-driven accretion.  The top
and bottom rows show snapshots on the radial-vertical and horizontal planes,
respectively, of the number density averaged over the region $-85\pc\leq
y\leq-35\pc$ (dotted lines in the bottom row) and $-50\pc\leq z \leq 50\pc$
(dotted lines in the top row) at $t/\torb=1.09$ (left), $1.12$ (center),
and $1.15$ (right).  The mean perturbed gravitational potential $\Delta \Phi
\equiv \Phi - \overline{\Phi}(z)$, where $\overline{\Phi}$ denotes the
horizontally-averaged potential, in units of $(\rm km\;s^{-1})^2$ is overlaid
as contours.  Cold cloudlets fall to the midplane, accrete surrounding gas, and
merge together to grow into a massive cloud that dominates the gravitational
potential.  \label{fig:acc}}
\end{figure}

Even without large-scale swing amplification, self-gravity (together with the
gravity of the stellar disk) plays an important role in creating ISM
structures.  Based on inspection of the evolving structure in our models, cold
cloudlets are seen to be drawn  together by gravity to create more massive
clouds.  This process is possible only because the ISM is a two-phase
cloud/intercloud medium.  The thermal pressure of cold cloudlets approximately
matches that of the surrounding warm medium at their surfaces, but the density
in the cold medium is two orders of magnitude higher than the warm medium.  As
a consequence, cold cloudlets move freely through the warm medium, falling
toward the midplane after reaching a maximum height.  Self-gravity then aids
their mutual collection to create a larger structure.  The successive snapshots
shown in Figure~\ref{fig:acc} illustrate  the formation of a massive cloud by
this process.

\begin{figure}
\epsscale{1.0}
\plotone{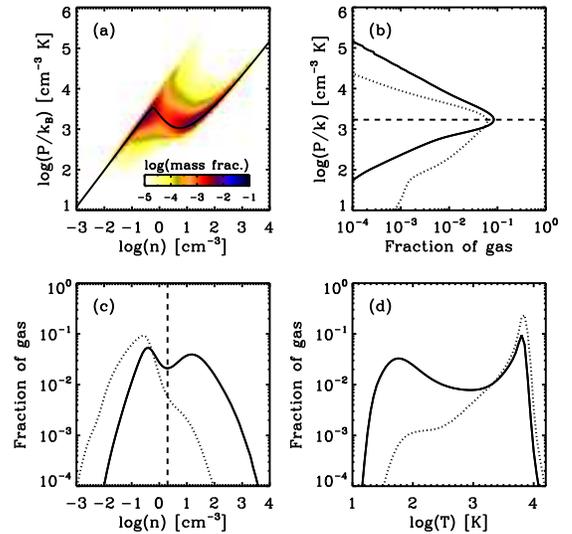}
\caption{ Probability density distributions of gas properties for the QA10-3DF
model. Time averages are taken after saturation, over $t/\torb=1-2$. (a)
Logarithmic mass fractions of QA10-3DF model in the $n$-$P$ plane.  The solid
curve indicates the locus of thermal equilibrium for the mean heating rate
$\abrackets{\Gamma}=0.61\Gamma_0$. (b-d) Mass-weighted (solid) and
volume-weighted (dotted) PDFs for (b) thermal pressure, (c) number density, and
(d) temperature.  The dashed lines in (b) and (c) denote the mean midplane
thermal pressure $\abrackets{\Pth/\kbol}=1.7\times10^3\pcc\Kel$ and number
density $\abrackets\rhomid=2.0\pcc$, respectively.  \label{fig:pdf}}
\end{figure}

Statistical distributions of the density, pressure, and temperature provide a
detailed picture of the thermodynamic state in our models.
Figure~\ref{fig:pdf} displays time-averaged probability distribution functions
(PDFs) of the QA10-3DF model, based on time averages over $t/\torb=1-2$. In
Figure~\ref{fig:pdf}(a), the mass fraction of gas as a function of number
density and thermal pressure is shown in logarithmic color scale.  The majority
of gas remains near the thermal equilibrium curve (solid line) defined by the
time-averaged heating rate $\abrackets{\Gamma}=0.61\Gamma_0$.
Figure~\ref{fig:pdf}(b-d) plots (b) thermal pressure, (c) number density, and
(d) temperature PDFs by mass (solid lines) and volume (dotted lines).  The
dashed lines in Figure~\ref{fig:pdf}(b) and (c) show the mean midplane thermal
pressure $\abrackets{\Pth/\kbol}=1.7\times10^3\pcc\Kel$ and the mean midplane
number density $\abrackets\rhomid=2.0\pcc$ for the QA10-3DF model.  The thermal
pressure lies mainly between the minimum value for cold gas in equilibrium,
$\Pmin$, and the maximum value for warm gas in equilibrium, $\Pmax$, with a
peak at the mean midplane pressure.  However, a significant amount of gas
($\sim 30-35\%$ by mass) has pressure higher and lower than $\Pmax$ and
$\Pmin$, respectively. The pressure distribution in our model is in part
similar to that of driven turbulence simulations in thermally bistable flows
\citep[e.g.][]{gaz05,gaz09,gaz13,sau13}.  Unlike those models, however, our
models also contain high-pressure cold gas confined by self-gravity, and
low-pressure warm gas found at high altitude.  Shock-heated warm gas at high
pressure is observed in both our models and in the non-self-gravitating,
unstratified simulations of other groups.

The density and temperature PDFs show bimodal distributions with cold, dense
and warm, rarefied phases as expected in the classical two-phase ISM
\citep[e.g.,][]{fie69,pio04}.  However, unlike the classical picture, the
distribution shows broadened peaks with a substantial fraction of gas
out-of-equilibrium.  These differences owe to several factors:  strong
turbulence that compresses and rarefies the gas continuously, SN events that
produce thermal transitions (in part induced by expanding strong shocks), and a
time-dependent heating rate such that the thermal equilibrium curve itself
fluctuates.  Since the level of turbulence in our models is transonic or
slightly subsonic for warm gas, the unimodal PDFs expected in highly supersonic
turbulent flows \citep{gaz05} are not found here.  Although velocity
dispersions are $\sim 20\%$ higher in the QA10-XZ model than in the 3D models,
we find that the PDFs for all QA10 models are very similar since the
differences in turbulence are small within one scale height ($|z|\simlt80\pc$)
where the bulk of mass is found (see Figure~\ref{fig:zprof}).

In all our simulations, disk evolution at later times ($t>\torb$) is similar to
that of the fiducial model. All models reach a quasi-steady state that includes
strong turbulence, cyclic formation and destruction of GBCs, vigorous stirring
of the population of small cloudlets, and quasi-periodic vertical disk
oscillations.  The vertical oscillations in our simulations are correlated over
the whole simulation domain. This is likely because the initial collapse of
non-turbulent, out-of-equilibrium gas to the midplane occurs simultaneously
over the whole region.\footnote{ Real galactic disks span $\sim$ 10 kpc in
radius, and any local vertical oscillations over scales of a few hundred pc
would be difficult to discern.  In addition, the presence of spiral arms, bars,
minor mergers, etc. may limit the development of local oscillations driven by
feedback from star formation.}  Vertical oscillations would be present in
reality,  but they would be correlated only over a smaller horizontal domain,
comparable to the size that is affected by feedback from a given star-forming
region.  For all model series, $\SigSFR$ and $\Ptot$ increase as the gas
surface density ($\Sigma$) and/or the depth of the gravitational potential well
(depending on $\Sigma$ and on the stellar density $\rhosd$) increase.  The
velocity dispersion $\vzdiff$ is more or less constant independent of model
parameters.  Detailed scaling relations and statistical properties will be
addressed below.

\begin{figure}
\epsscale{1.0}
\plotone{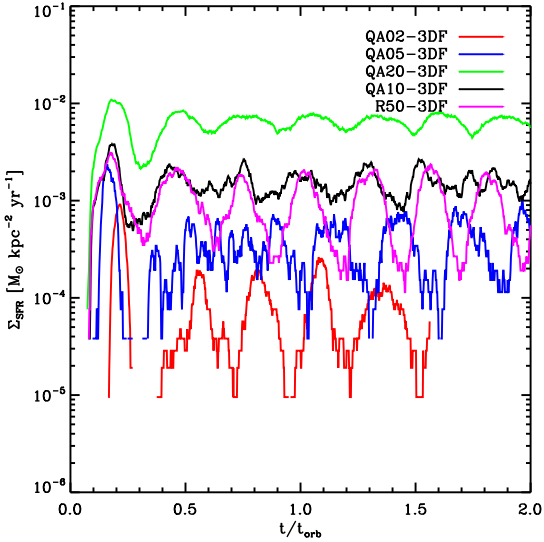}
\caption{ Time evolution of the SFR surface density $\SigSFR$ for all 3DF
models.  The amplitude of temporal fluctuations increases as $\SigSFR$
decreases, because the system is more stochastic.  \label{fig:sfr}}
\end{figure}

For models with low gas and stellar densities, the star formation events become
rare and stochastic, leading to large amplitude fluctuations in $\SigSFR$. In
high-$\frad$ models,  the heating rate is highly sensitive to $\SigSFR$, again
resulting in large temporal variation in ISM properties and $\SigSFR$.  The XZ
models in Paper~I were unable to fully address the effect of strong temporal
fluctuation of the SFR in low-$\SigSFR$ and high-$\frad$ models because it was
necessary to adopt $\tbin$ much longer than the realistic lifetime of massive
stars (cf. Figure~\ref{fig:hst1}(b)). In the present 3DF models with large
azimuthal domains, we can adopt $\tbin$ close to the FUV luminosity-weighted
life time $t_{\rm FUV}\sim10\Myr$.  Figure~\ref{fig:sfr} plots time evolution
of $\SigSFR$ for all 3DF models.  Relative to $\torb$, the fluctuation periods
of $\SigSFR$ are similar for all the models  shown  since the ratio  $t_{\rm
osc}/\torb \approx 0.25\Omega/(4G\rhosd)^{1/2}=0.27$ is constant.  The
low-density and high-$\frad$ models show order-of-magnitude fluctuations in
$\SigSFR$, while fluctuations are only at a factor of 2 -- 3 level for models
at higher density and $\frad=1$.  In the high-$\frad$ model, the large
fluctuations of the SFR are self-reinforcing because fluctuations of the
heating rate follow the SFR as $\Gamma\propto\frad\SigSFR$ and the resulting
fluctuations in cold gas content lead to varying $\SigSFR$.

\begin{figure}
\epsscale{1.0}
\plotone{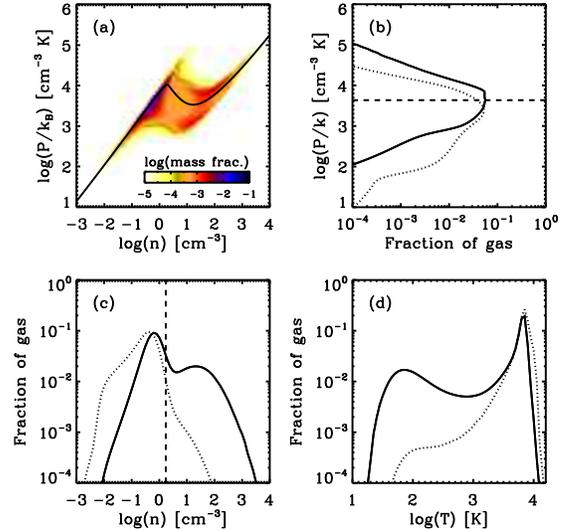}
\caption{ The same as Figure~\ref{fig:pdf}, but for model R50-3DF with high
radiation feedback efficiency $\frad=5.0$.  Comparing panel (a) to
Figure~\ref{fig:pdf}(a), signatures of a ``high'' and a ``low'' state are
evident in this time-averaged distribution.  For these states, the thermal
equilibrium curve lies either above (``high'' state) or below (``low'' state)
the thermal equilibrium curve at the mean heating rate of
$\abrackets{\Gamma}=1.9\Gamma_0$ (heavy curve in panel (a)).  The mean midplane
thermal pressure is $\abrackets{\Pth/\kbol}=4.3\times10^3\pcc\Kel$, and the
mean midplane number density is $\abrackets{\rhomid}=1.7\pcc$, as indicated in
panels (b) and (c).  \label{fig:pdf_r50}}
\end{figure}

Figure~\ref{fig:pdf_r50} displays (a) the gas mass fraction in the
density-pressure phase plane, and (b) thermal pressure, (c) number density, and
(d) temperature PDFs of the R50-3DF enhanced-heating model with $\frad=5.0$.
While the ranges of pressure, density, and temperature are similar to the QA10
$\frad=1$ model shown in Figure \ref{fig:pdf}, this high-$\frad$ model shows
higher mean thermal pressure, a lower fraction of cold, dense gas, and a
broader peak in the pressure PDF.  These differences are a consequence of both
the higher heating rate and increased stochasticity of this model.  For the
R50-3DF model, the time evolution of the SFR in Figure~\ref{fig:sfr} shows two
distinct levels of $\SigSFR$ rather than the moderate fluctuations about a mean
value seen for model QA10-3DF.  Since the gas is approximately in instantaneous
thermal equilibrium at either the high or low $\SigSFR$ state, with either a
high or low heating rate and the corresponding equilibrium curve, traces of the
two states are evident in the distributions shown in Figure~\ref{fig:pdf_r50}a.
As a consequence of the well-separated high and low states, the time-averaged
pressure PDF is quite broad.  In spite of the two-state behavior of the R50-3DF
model, the mean  values of $\SigSFR$, $\Gamma$, $\Pth$, and $\rhomid$ are quite
similar (see Table \ref{tbl:stat1}) to those in the R50-3DS and R50-XZ models,
in which fluctuations are reduced by adopting $\tbin =0.5\torb>t_{\rm osc}$ for
computing the heating rate.

Detailed examination of our low-density models (QA02-3DF and QA05-3DF) shows
that stochasticity similarly leads to broadened distributions of $\Pth$
compared to the fiducial model.  In real galaxies, regions with low gas and
stellar densities also tend to have lower metallicity, such that $\frad$
increases, which could increase fluctuations even more compared to these models
with $\frad=1$.  Even in highly stochastic 3DF models, however, we find that
the mean saturated-state properties are similar to those found from XZ models,
confirming the results of Paper~I.

\begin{deluxetable*}{lccccc} \tablecolumns{6}
\tabletypesize{\scriptsize} \tablewidth{0pt}
\tablecaption{Disk Properties 1\label{tbl:stat1}}
\tablehead{
\colhead{Model} &
\colhead{$\log{\SigSFR}$} &
\colhead{$\log{\Pth/\kbol}$} &
\colhead{$\log{\Pturb/\kbol}$} &
\colhead{${\rhomid}$} &
\colhead{${\Hdiff}$} \\
\colhead{(1)} & \colhead{(2)} & \colhead{(3)} &
\colhead{(4)} & \colhead{(5)} & \colhead{(6)}
}
\startdata
  QA02-3DF & $-4.11\pm 0.32 $ & $ 2.23\pm 0.16 $ & $ 2.52\pm 0.35 $ & $ 0.19\pm 0.13 $ & $ 306\pm 111 $ \\
  QA02-3DS & $-4.04\pm 0.09 $ & $ 2.29\pm 0.23 $ & $ 2.56\pm 0.54 $ & $ 0.14\pm 0.11 $ & $ 337\pm 100 $ \\
   QA02-XZ & $-4.20\pm 0.23 $ & $ 1.94\pm 0.35 $ & $ 2.61\pm 1.51 $ & $ 0.05\pm 0.08 $ & $ 342\pm 111 $ \\
\hline
  QA05-3DF & $-3.43\pm 0.27 $ & $ 2.70\pm 0.15 $ & $ 3.19\pm 0.35 $ & $ 0.53\pm 0.34 $ & $ 160\pm  53 $ \\
  QA05-3DS & $-3.43\pm 0.06 $ & $ 2.69\pm 0.19 $ & $ 3.19\pm 0.38 $ & $ 0.35\pm 0.24 $ & $ 170\pm  29 $ \\
   QA05-XZ & $-3.52\pm 0.12 $ & $ 2.53\pm 0.22 $ & $ 3.03\pm 0.72 $ & $ 0.39\pm 0.30 $ & $ 174\pm  37 $ \\
\hline
  QA10-3DF & $-2.82\pm 0.12 $ & $ 3.23\pm 0.10 $ & $ 3.69\pm 0.15 $ & $ 1.46\pm 0.40 $ & $  85\pm  13 $ \\
  QA10-3DS & $-2.84\pm 0.03 $ & $ 3.25\pm 0.11 $ & $ 3.86\pm 0.42 $ & $ 1.47\pm 0.55 $ & $  84\pm  12 $ \\
   QA10-XZ & $-2.74\pm 0.11 $ & $ 3.24\pm 0.15 $ & $ 3.85\pm 0.60 $ & $ 1.12\pm 0.58 $ & $  92\pm  18 $ \\
\hline
  QA20-3DF & $-2.18\pm 0.06 $ & $ 3.79\pm 0.04 $ & $ 4.20\pm 0.09 $ & $ 3.43\pm 0.58 $ & $  47\pm   4 $ \\
  QA20-3DS & $-2.20\pm 0.02 $ & $ 3.86\pm 0.06 $ & $ 4.32\pm 0.23 $ & $ 3.83\pm 0.80 $ & $  47\pm   5 $ \\
   QA20-XZ & $-2.06\pm 0.10 $ & $ 3.86\pm 0.07 $ & $ 4.19\pm 0.63 $ & $ 2.76\pm 0.70 $ & $  51\pm   5 $ \\
\hline
   R50-3DF & $-3.02\pm 0.30 $ & $ 3.64\pm 0.19 $ & $ 3.60\pm 0.49 $ & $ 1.31\pm 0.88 $ & $  99\pm  29 $ \\
   R50-3DS & $-3.05\pm 0.05 $ & $ 3.72\pm 0.09 $ & $ 3.82\pm 0.53 $ & $ 1.34\pm 0.46 $ & $  92\pm  13 $ \\
    R50-XZ & $-2.96\pm 0.22 $ & $ 3.69\pm 0.16 $ & $ 3.31\pm 0.37 $ & $ 1.29\pm 0.58 $ & $  97\pm  20 $
\enddata
\tablecomments{
The temporal averages and standard deviations are taken over $t/\torb=1-2$ for
3D models and $t/\torb=2-3$ for XZ models.  Col. (2): Logarithm of the SFR
surface density (M$_\odot\kpc^{-2}\yr^{-1}$).  Cols.  (3)-(4): Logarithm of the
midplane thermal and turbulent pressures over $\kbol$ (cm$^{-3}\;K$).  Col.
(5): Midplane number density of hydrogen (cm$^{-3}$).  Col. (6): Scale height
of the diffuse component (pc).  See Section \ref{sec:stat} for definitions.  }
\end{deluxetable*}

\begin{deluxetable*}{lcccccc}
\tabletypesize{\scriptsize} \tablewidth{0pt}
\tablecaption{Disk Properties 2\label{tbl:stat2}}
\tablehead{
\colhead{Model} &
\colhead{${\vzdiff}$} &
\colhead{${\vthdiff}$} &
\colhead{$\alpha_v$} &
\colhead{$\alpha_P$} &
\colhead{${\fdiff}$} &
\colhead{$\tilde{f}_w$} \\
\colhead{(1)} & \colhead{(2)} & \colhead{(3)} &
\colhead{(4)} & \colhead{(5)} & \colhead{(6)} & \colhead{(7)} }
\startdata
  QA02-3DF & $ 4.33\pm 1.07 $ & $ 4.07\pm 0.57 $ & $ 2.13\pm 0.64 $ & $ 2.95\pm 1.70 $ & $ 0.99\pm 0.01 $ & $ 0.21\pm 0.17 $ \\
  QA02-3DS & $ 4.73\pm 1.75 $ & $ 4.22\pm 0.62 $ & $ 2.26\pm 1.00 $ & $ 2.86\pm 2.49 $ & $ 0.99\pm 0.02 $ & $ 0.34\pm 0.14 $ \\
   QA02-XZ & $ 6.20\pm 4.57 $ & $ 3.27\pm 0.78 $ & $ 4.59\pm 5.56 $ & $ 5.75\pm16.94 $ & $ 0.92\pm 0.07 $ & $ 0.23\pm 0.12 $ \\
\hline
  QA05-3DF & $ 5.03\pm 1.00 $ & $ 3.85\pm 0.78 $ & $ 2.71\pm 0.97 $ & $ 4.09\pm 2.73 $ & $ 0.95\pm 0.03 $ & $ 0.26\pm 0.16 $ \\
  QA05-3DS & $ 5.11\pm 1.43 $ & $ 3.90\pm 0.26 $ & $ 2.72\pm 0.99 $ & $ 4.09\pm 3.02 $ & $ 0.96\pm 0.03 $ & $ 0.30\pm 0.05 $ \\
   QA05-XZ & $ 6.28\pm 2.95 $ & $ 3.25\pm 0.43 $ & $ 4.74\pm 3.65 $ & $ 4.16\pm 5.46 $ & $ 0.90\pm 0.06 $ & $ 0.25\pm 0.07 $ \\
\hline
  QA10-3DF & $ 4.67\pm 0.47 $ & $ 4.11\pm 0.26 $ & $ 2.29\pm 0.31 $ & $ 3.88\pm 1.21 $ & $ 0.88\pm 0.03 $ & $ 0.34\pm 0.06 $ \\
  QA10-3DS & $ 5.11\pm 1.01 $ & $ 3.88\pm 0.23 $ & $ 2.73\pm 0.71 $ & $ 5.08\pm 4.06 $ & $ 0.87\pm 0.05 $ & $ 0.32\pm 0.05 $ \\
   QA10-XZ & $ 7.23\pm 2.28 $ & $ 3.73\pm 0.42 $ & $ 4.75\pm 2.51 $ & $ 5.02\pm 5.73 $ & $ 0.77\pm 0.09 $ & $ 0.32\pm 0.07 $ \\
\hline
  QA20-3DF & $ 5.05\pm 0.32 $ & $ 4.47\pm 0.22 $ & $ 2.27\pm 0.20 $ & $ 3.53\pm 0.58 $ & $ 0.72\pm 0.03 $ & $ 0.41\pm 0.05 $ \\
  QA20-3DS & $ 5.71\pm 0.68 $ & $ 4.35\pm 0.18 $ & $ 2.72\pm 0.43 $ & $ 3.86\pm 1.55 $ & $ 0.71\pm 0.04 $ & $ 0.38\pm 0.04 $ \\
   QA20-XZ & $ 6.94\pm 1.67 $ & $ 4.59\pm 0.22 $ & $ 3.29\pm 1.12 $ & $ 3.17\pm 3.16 $ & $ 0.54\pm 0.06 $ & $ 0.46\pm 0.05 $ \\
\hline
   R50-3DF & $ 5.51\pm 1.18 $ & $ 5.35\pm 0.93 $ & $ 2.06\pm 0.59 $ & $ 1.92\pm 1.12 $ & $ 0.91\pm 0.07 $ & $ 0.61\pm 0.26 $ \\
   R50-3DS & $ 4.48\pm 1.18 $ & $ 5.77\pm 0.21 $ & $ 1.60\pm 0.32 $ & $ 2.25\pm 1.54 $ & $ 0.91\pm 0.03 $ & $ 0.74\pm 0.06 $ \\
    R50-XZ & $ 4.49\pm 1.85 $ & $ 5.70\pm 0.49 $ & $ 1.62\pm 0.52 $ & $ 1.42\pm 0.39 $ & $ 0.85\pm 0.07 $ & $ 0.73\pm 0.12 $
\enddata
\tablecomments{
The temporal averages and standard deviations are taken over $t/\torb=1-2$ for
3D models and $t/\torb=2-3$ for XZ models.  Cols. (2)-(3): Vertical turbulent
and thermal velocity dispersions of the diffuse gas (${\rm km\;s^{-1}}$). Cols.
(4)-(5): Ratios of total pressure to thermal pressure calculated from the
mass-weighted velocity dispersions ($\alpha_v$) and the midplane pressures
($\alpha_P$).  Cols. (6)-(7): the mass fraction of the diffuse gas ($\fdiff$),
and the square of mass-weighted thermal to warm-medium thermal speed
($\vthdiff^2/c_w^2=\tilde{f}_w$) in the diffuse gas.  See Section
\ref{sec:stat} for definitions.}
\end{deluxetable*}

\subsection{Statistical Properties}\label{sec:stat}

\begin{figure}
\epsscale{1.0}
\plotone{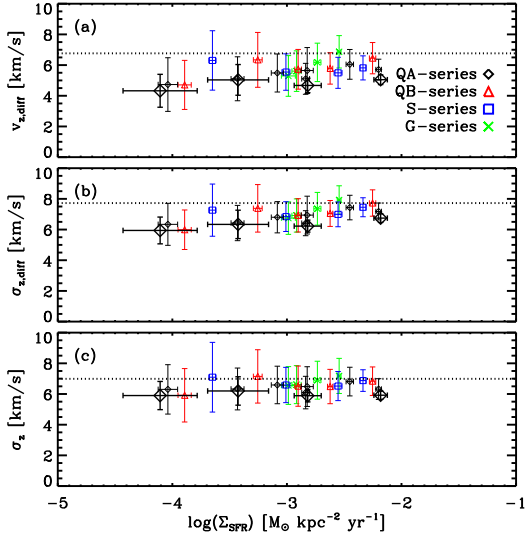}
\caption{
(a) The vertical turbulent velocity dispersion of the diffuse gas $\vzdiff$,
(b) the total (turbulent+thermal) velocity dispersion of the diffuse gas
$\szdiff$, and (c) the total velocity dispersion of all the gas (diffuse + GBC)
$\sz$, as functions of the SFR surface density $\SigSFR$.  Results for all 3D
models except the R-series are shown; thick large symbols denote the results
from 3DF models, while thin small symbols denote the results from 3DS models.
The points and errorbars give the mean values and standard deviations over
$t/\torb=1-2$.  Mean values from the 3D models are $\vzdiff=5.6\pm0.6 \kms$,
$\szdiff=6.9\pm0.5\kms$, and $\sz=6.5\pm0.4\kms$.  The horizontal dotted lines
indicate the mean values ($\vzdiff=6.8\pm0.6 \kms$, $\szdiff=7.7\pm0.6\kms$,
and $\sz=7.0\pm0.4\kms$) from all XZ models (Paper~I) except the R-series.  Due
to geometric effects, velocity dispersions from the XZ models are slightly
larger than those from the 3D models.  \label{fig:v_sfr}}
\end{figure}

\begin{figure}
\epsscale{1.0}
\plotone{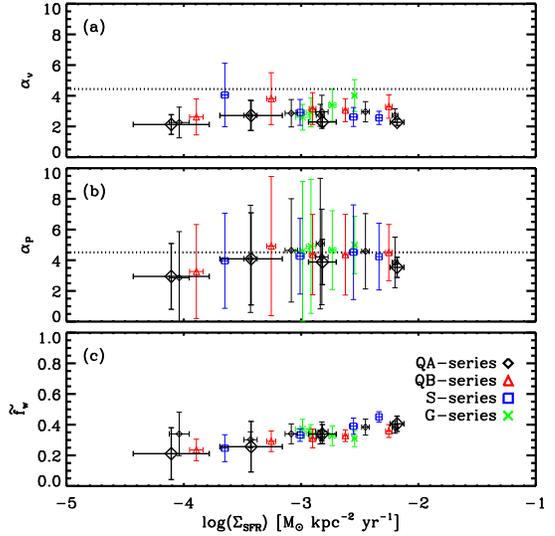}
\caption{ (a)
The ratio of total-to-thermal velocity dispersion for the diffuse gas
$\alpha_v\equiv1+\vzdiff^2/\vthdiff^2$ for velocity dispersion weighted by mass
over the whole domain; (b) the ratio of total-to-thermal pressure for gas at
the midplane $\alpha_P\equiv 1+\Pturb/\Pth$, and (c) the ratio of mass-weighted
thermal to warm-medium thermal speed $\vthdiff^2/c_w^2=\tilde{f}_w$.  Results
for all 3D models except the R series are shown.  Symbols have the same
meanings as in Figure~\ref{fig:v_sfr}.  The dotted lines in (a) and (b) are the
mean values of all XZ models (Paper~I), $\alpha_v=4.4$ and $\alpha_P=4.5$. For
3D models, $\alpha_P$ is consistent with the value in  XZ models, but
$\alpha_v$ is smaller (see text).  The quantity $\tilde f_w$ (essentially the
warm mass fraction) ranges over $0.2-0.5$ as in Paper~I.  \label{fig:afw}}
\end{figure}

All our models reach a quasi-steady state after two or three vertical
oscillation times (see Figure \ref{fig:sfr}), which is $<\torb$ for our model
parameters.  We thus investigate statistical properties of 3D models by
averaging over $t/\torb=1-2$.  Tables~\ref{tbl:stat1} and \ref{tbl:stat2} list
the mean values and standard deviations of key physical quantities used in
Paper~I to test the thermal/dynamical equilibrium model of OML10 and OS11. In
the Tables, we present the results for all 3DF models together with the
corresponding 3DS and XZ counterparts. In Figures \ref{fig:v_sfr} -
\ref{fig:sfr_p}, we also include results from additional 3DS models without 3DF
counterparts.  Hereafter, in reporting properties of the models, we use time
averages over $t/\torb=1-2$ for 3D models and $t/\torb=2-3$ for XZ models (as
in Paper~I), unless stated otherwise. Angle brackets will be omitted for
convenience.

Column (1) of Tables~\ref{tbl:stat1} and \ref{tbl:stat2} gives the name of each
model, consisting of the model name listed in Table~\ref{tbl:model} together
with suffixes of 3DF, 3DS, and XZ to indicate the simulation domain size and
geometry.  In Table~\ref{tbl:stat1}, Column (2) lists $\log\SigSFR$ in units of
$\sfrunit$. We list in Columns (3) and (4) $\log(\Pth/\kbol)$ and
$\log(\Pturb/\kbol)$, respectively, in units of $\Punit$. Column (5) gives the
midplane hydrogen number density $\rhomid$ of the diffuse component in units of
$\pcc$, defined in the same way with equation (\ref{eq:Pth}) but for number
density $n$ rather than thermal pressure $P$ in the integrand.  Similarly, the
scale height of the diffuse gas $\Hdiff\equiv[\int\rho z^2\Theta(n\!<\!\ngbc)
dV/\int\rho\Theta(n\!<\!\ngbc) dV]^{1/2}$ is listed in Column (6) in units of
$\pc$.

In Table~\ref{tbl:stat2}, Columns (2) and (3) list respectively $\vzdiff$ and
$\vthdiff$ in units of $\kms$.  In Columns (4) and (5), we list respectively
the ratios of total to thermal pressure defined in two different ways,
$\alpha_v\equiv(\vthdiff^2+\vzdiff^2)/\vthdiff^2$ and
$\alpha_P\equiv(\Pth+\Pturb)/\Pth$.  Here, $\alpha_v$ averages over the whole
volume, while $\alpha_P$ averages only at the midplane.  Column (6) gives
$\fdiff$, the mass fraction of the diffuse component ($n<\ngbc$), and Column
(7) gives $\tilde{f}_w\equiv\vthdiff^2/c_w^2$, the ratio of the mass-weighted
sound speed to the thermal velocity of warm gas (this is essentially equal to
the warm gas fraction of mass in the diffuse component); these parameters are
used in the OML10 theory.

Figure~\ref{fig:v_sfr} plots the mean values of (a) $\vzdiff$,  (b)
$\szdiff\equiv(\vzdiff^2+\vthdiff^2)^{1/2}$, and (c) the velocity dispersion of
whole medium $\sz\sim\fdiff^{1/2}\szdiff$, as functions of $\SigSFR$ for all 3D
models except the R-series. The dotted horizontal line in each panel denotes
the mean value from all XZ models (the models of Paper~I).  The mean values of
all 3D models give $\vzdiff=5.6\pm0.6\kms$, $\szdiff=6.9\pm0.5\kms$, and
$\sz=6.5\pm0.4\kms$.  As in Paper~I, we find that the velocity dispersion is
more-or-less constant over two orders of magnitude in $\SigSFR$. Due to
geometrical effects in the expansion of SN remnants (see  Section
\ref{sec:evol}), turbulent velocity dispersions are slightly smaller (about
$18\%$) in 3D models than in XZ models.

Figure~\ref{fig:afw} plots the mean values of (a) $\alpha_v$, (b) $\alpha_P$,
and (c) $\tilde f_w$ as functions of $\SigSFR$ for all 3D models except the R
series.  The quantities $\alpha_v$ and $\alpha_P$ indicate whether the velocity
dispersion and pressure in the ISM is thermally ($\alpha <2$) or dynamically
($\alpha > 2$) dominated.    The dotted lines in (a) and (b) indicate the mean
values of $\alpha_v=4.4$ and $\alpha_P=4.5$ found from XZ models in Paper~I.
In XZ models, $\alpha_v\sim\alpha_P$ even though both $v_{\rm th}(z)$ and
$v_z(z)$ vary significantly along the vertical direction since the profiles
have similar shape (see blue lines in Figure~\ref{fig:zprof}).  In contrast,
the vertical profiles of thermal and turbulent velocity dispersions have
different shapes in 3D models (see black and red lines in
Figure~\ref{fig:zprof}). This results in differences between $\alpha_v$ and
$\alpha_P$.  Averaging over all models except the R series yields
$\alpha_v=2.9$ and $\alpha_P=4.2$, showing that the turbulent component of the
mass-weighted velocity dispersion underestimates the midplane value.  In the
remaining sections, we will use $\alpha_P$ as a representative value, rather
than $\alpha_v$.  Figure \ref{fig:afw}(c) shows that the warm mass fraction
$\tilde f_w$  remains in the range $0.2$ to $0.5$ for all models,
as previously found in Paper~I. The fiducial value $\tilde f_w \sim 0.5$
adopted in OML10 is consistent with the range of our simulations.

\subsection{Comparison with Thermal/Dynamical Equilibrium Model}\label{sec:equil}

It is natural to expect that a disk system that reaches a quasi-steady state
satisfies vertical dynamical equilibrium in an average sense.  That is, the
diffuse ISM must maintain force balance between upward pressure forces (thermal
and turbulent) and downward gravitational forces (arising from the diffuse gas
itself as well as stars, dark matter, and GBCs) (\citealt{pio07,koy09b}; OML10,
Paper~I). By taking time and horizontal averages, the vertical momentum
equation in equilibrium can be written (OML10, Paper~I) as $\Ptot= \PtotDE$ for
\begin{eqnarray}\label{eq:PDE}
&\PtotDE&\equiv\fdiff\frac{\pi G\Sigma^2}{4}\times\nonumber\\
&&\cbrackets{(2-\fdiff)+ \sbrackets{\rbrackets{2-\fdiff}^2+
\frac{32\szdiff^2\rhosd}{\pi^2 G\Sigma^2}}^{1/2}},
\end{eqnarray}
where $\PtotDE$ represents the weight of the diffuse ISM.\footnote{In
observational estimates, it is often assumed that all of the gas is in vertical
equilibrium.  Equation (\ref{eq:PDE}) instead represents the case in which all
of the gas contributes to the vertical gravity, but only the diffuse gas (a
fraction $\fdiff$ of the total) is in vertical equilibrium.  For outer disks
(as studied here) most gas is diffuse ($\fdiff \sim 1$), but for inner disks an
increasing fraction of the mass 
may be in GBCs that are at higher pressure than their
surroundings.} Note that in OML10, $c_w^2\tilde f_w \alpha$ was used instead of
$\szdiff^2$, but here we use the latter as  the total vertical velocity
dispersion $\szdiff$ is directly measurable in our simulations. Since in
\S\ref{sec:stat} we explicitly measure $\fdiff$ and $\szdiff$ as well as
$\Ptot$ for given input parameters $(\Sigma, \rhosd)$, a direct comparison
between $\PtotDE$ and $\Ptot$ is possible.

\begin{figure}
\epsscale{1.0}
\plotone{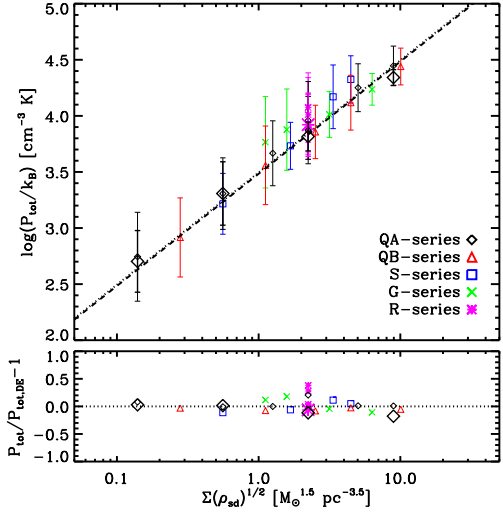}
\caption{ Top: total pressure of the diffuse gas measured in the simulations as
a function of $\Sigma\sqrt{\rhosd}$.  The symbols denote the same meanings as
in Figure~\ref{fig:v_sfr}.  The dashed and dotted lines show our best fit for
all 3D models and XZ models, respectively.  The good agreement between the two
confirms the reliability of XZ models from Paper~I.  Bottom: relative
differences between the measured midplane pressure $\Ptot$ and the pressure
predicted for dynamical equilibrium $\PtotDE$ using Equation (\ref{eq:PDE}).
The mean relative difference is only 12\%, showing that vertical dynamical
equilibrium indeed holds in an average sense.  \label{fig:ptot}}
\end{figure}

Figure~\ref{fig:ptot} gives {\it measured} total pressure $\Ptot$ as a function
of $\Sigma\sqrt{\rhosd}$ for all 3D models.  The relative difference between
{\it measured} and {\it predicted} (dynamical equilibrium) total pressure,
$\Ptot/\PtotDE-1$, is shown in the bottom panel.  The mean relative difference
is only $12\%$, showing that vertical dynamical equilibrium indeed holds.  For
the models we consider here, in which the external gravity exceeds the
self-gravity of the diffuse gas, the last term in the square root of Equation
(\ref{eq:PDE}) is dominant, and we expect:
\begin{eqnarray}\label{eq:PDEapprox}
\PtotDE &\approx& \fdiff \szdiff\Sigma(2G\rhosd)^{1/2}.
\end{eqnarray}
If $\fdiff\szdiff$ is insensitive to model parameters (as in Paper~I) and
$\Ptot \approx \PtotDE$, then the midplane total pressure is expected to
correlate with $\Sigma\sqrt{\rhosd}$.  This correlation is evident in the
dashed line in Figure~\ref{fig:ptot}; the best fit coefficient for the 3D model
results yields:
\begin{eqnarray}\label{eq:Ptotfit}
\Ptot& =& 9.6\times10^3\kbol\Punit\times\nonumber\\
&&\rbrackets{\frac{\Sigma}{10\Surf}}
\rbrackets{\frac{\rhosd}{0.1\Msun\pc^{-3}}}^{1/2}.
\end{eqnarray}
In Figure \ref{fig:ptot}, we overplot as a dotted line the best fit for the XZ
models (eq.~(37) in Paper~I), which agrees very well with the dashed line and
the 3D model results.

In addition to the force balance between total pressure and gravity, individual
balances between thermal cooling and heating, and between turbulence driving
and dissipation, are expected for a quasi-steady state. The gas  is heated by
FUV radiation from newly formed massive stars, with a  heating rate that varies
in proportion to the SFR (see Equation~(\ref{eq:heat})).  Owing to the very
short cooling time compared to the dynamical time, the gas generally resides
near thermal equilibrium, although a non-negligible fraction deviates from
thermal equilibrium due to strong turbulence and temporal heating fluctuations
(see Figures~\ref{fig:pdf} and \ref{fig:pdf_r50}). Near the midplane, where
GBCs and stars form, a cold phase should coexist with warm phase, such that the
midplane thermal pressure $\Pth$ should lie between $\Pmin$ and $\Pmax$.
Figures~\ref{fig:pdf} and \ref{fig:pdf_r50} indeed show that the pressure lies
in the neighborhood defined by $\Pmin$ and $\Pmax$ for the mean heating rate.

\begin{figure}
\epsscale{1.0}
\plotone{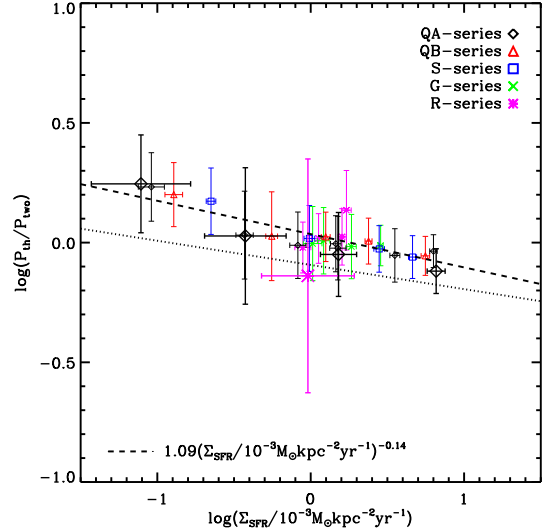}
\caption{
Measured midplane thermal pressure of the diffuse gas $\Pth$ relative to the
two-phase thermal equilibrium pressure $\Ptwo$ as a function of $\SigSFR$.  The
symbols have the same meanings as in Figure~\ref{fig:v_sfr}.  The dashed and
dotted lines give the best fits for 3D and XZ models, respectively.  The result
from our simulations that $\Pth\sim\Ptwo \propto \SigSFR$ over more than two
orders of magnitude in $\SigSFR$ implies that the radiative heating from star
formation is balanced by cooling within a local (vertical) dynamical time.
\label{fig:fth}}
\end{figure}

In OML10, the mean value of $\Pth$ at the midplane is assumed to approach the
geometric-mean pressure $\Ptwo\equiv(\Pmin\Pmax)^{1/2}$.  The adopted cooling
function and heating rate given by equations (\ref{eq:cool}) and
(\ref{eq:heat}) yield $\Ptwo/\kbol=3.1\times10^3\pcc\Kel
(\frad\SigSFR/\SigSFRsun)$, assuming that the contribution from the
metagalactic FUV radiation is negligible. Our measurements for the mean value
of $\SigSFR$ from the numerical simulations give the mean values of $\Ptwo$ for
each model, which can be directly compared to the values of the measured mean
midplane thermal pressure $\Pth$.  Figure~\ref{fig:fth} plots $\Pth/\Ptwo$ as a
function of $\SigSFR$ for all 3D models. The errorbars denote the standard
deviations of temporal fluctuation, which amount to $\sim 0.2-0.3{\rm\; dex}$ .
Our best fit for 3D results,
\begin{equation}\label{eq:fth}
\frac{\Pth}{\Ptwo}=1.09\rbrackets{\frac{\SigSFR}{10^{-3}\sfrunit}}^{-0.14},
\end{equation}
is shown as the dashed line, while the dotted line denotes the best fit for XZ
models from Paper~I.  We conclude that $\Pth$ is indeed comparable to $\Ptwo$,
over two orders of magnitude variation in $\SigSFR$.  Our results show that
$\Pth\sim\Ptwo$ within $60\%$, supporting the approximation adopted in the
analytic theory of OML10.  The mean values of the midplane thermal pressure in
3D models are slightly higher ($\sim 20\%$) than that of the XZ models, but the
difference is smaller than the temporal fluctuations.  Significant warm gas in
the R50-3DF model (big magenta asterisk) puts it at lower average pressure than
the mean relation; as discussed in section \ref{sec:evol}, there are also large
fluctuations in this model.

Dynamical energy injection through SN feedback connects the turbulent pressure
and the SFR, similar to the connection between thermal pressure and the SFR
from UV heating.  Since the turbulent velocity dispersion does not evolve
secularly (see Figure \ref{fig:hst2}), the rate of turbulence driving must
balance turbulent dissipation.  The rate of turbulent driving per unit area per
unit mass is expected to be $\propto(\Psn/\Msn)\SigSFR$, where $\Psn$ is the
mean radial momentum injected to the ISM by an expanding SN remnant, and $\Msn$
is the total mass in stars per SN (averaged over the stellar mass function).
For spherical blasts centered on the midplane, the vertical momentum injection
rate per unit area to each side of the disk is $\Pdriv=0.25(\Psn/\Msn)\SigSFR$
(Paper~I, OS11, SO12).  The turbulent momentum flux in the vertical direction
through the disk can be expressed as $\Pturb \equiv f_p\Pdriv$, where
$f_p\sim1$ if dissipation balances driving within a dynamical time scale.

\begin{figure}
\epsscale{1.0}
\plotone{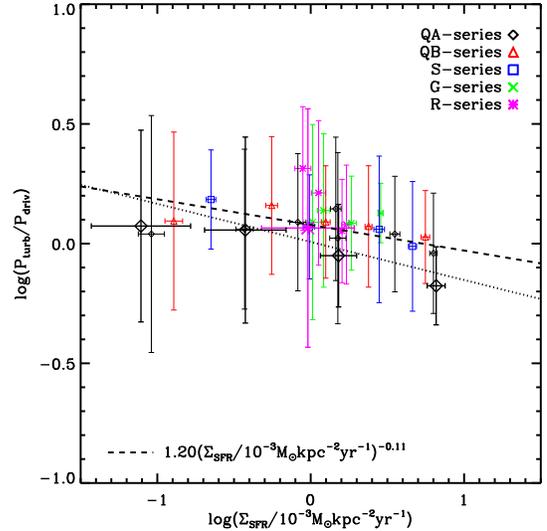}
\caption{
Measured midplane turbulent pressure of the diffuse gas $\Pturb$ relative to
the characteristic vertical momentum flux injected by star formation feedback
$\Pdriv$ (see text), as a function of $\SigSFR$.  The symbols have the same
meanings as in Figure~\ref{fig:v_sfr}.  Dashed and dotted lines give the best
fits for 3D and XZ models, respectively.  The result that $\Pturb\sim\Pdriv
\propto\SigSFR$ from our simulations, over more than two orders of magnitude in
$\SigSFR$, implies that the turbulent energy driving from star formation is
balanced by dissipation within a local (vertical) dynamical time.
\label{fig:fp}}
\end{figure}

Using the adopted value $\Psn/\Msn=3000\kms$, the fiducial momentum injection
rate varies with the star formation rate as $\Pdriv/\kbol=3.6\times10^3\Punit
(\SigSFR/10^{-3} \sfrunit)$. We make a direct comparison between the measured
midplane turbulent pressure $\Pturb$ and $\Pdriv$. Figure~\ref{fig:fp} plots
the mean values of $f_p\equiv \Pturb/\Pdriv$ as a function of $\SigSFR$ for all
3D models.  The errorbars denote the standard deviations of temporal
fluctuations, which amount to $\sim 0.3-0.6{\rm\; dex}$.  The dashed line is
our best fit omitting the R-series,
\begin{equation}\label{eq:fp}
\frac{\Pturb}{\Pdriv}=1.20\rbrackets{\frac{\SigSFR}{10^{-3}\sfrunit}}^{-0.11},
\end{equation}
while the fit for the XZ models (Paper~I) is overplotted as dotted line with
slightly steeper slope of $-0.17$. The results from the 3D models are in
overall good agreement with our previous XZ models of Paper~I, confirming that
$f_p$ is order-unity and approximately constant for a wide range of disk
conditions and star formation rates (see also SO12, which shows that $f_p$ is
similar in the starburst regime).

Since $\Pth\sim\Ptwo$ and $\Pturb\sim\Pdriv$, both thermal and turbulent
pressures at the midplane are nearly linearly proportional to $\SigSFR$.
Following Paper~I, we define
\begin{eqnarray}
\frac{P_{\rm th}/k_B}{10^3 \Punit}&\equiv & \etath
\frac{\SigSFR}{10^{-3} \sfrunit},
\label{eq:etath}
\\
\frac{P_{\rm turb}/k_B}{10^3 \Punit}&\equiv& \etaturb
\frac{\SigSFR}{10^{-3} \sfrunit}.
\label{eq:etaturb}
\end{eqnarray}
such that the yield coefficients $\etath$ and $\etaturb$ measure the thermal
and turbulent efficacies of feedback, respectively.  In the analytic model for
thermal/dynamical equilibrium (OML10; OS11), $\etath= 1.2\frad$ where $\frad=
[0.25+0.75 Z_d'(\Sigma/10 \Msun \pc^{-2})^{0.4}]^{-1}$ using the
heating/cooling model of \citet{wol03}, and $\etaturb= 3.6 f_p$ for
$\Psn/\Msn=3000\kms$.  Our direct measurements fitting the simulation results
(equations (\ref{eq:fth}) and (\ref{eq:fp})) give
\begin{eqnarray}
\etath= 1.3\frad\left(\frac{\SigSFR}{10^{-3}\sfrunit}\right)^{-0.14},
\label{eq:etathnum}\\
\etaturb= 4.3
\left(\frac{\SigSFR}{10^{-3}\sfrunit}\right)^{-0.11}.
\label{eq:etaturbnum}
\end{eqnarray}
These results for yields are very close to those adopted in the analytic model,
and show quite weak dependence on the SFR.  For Solar neighborhood conditions,
the yield coefficients are $\etath=1.1$ and $\etaturb=3.9$, remarkably close to
the adopted values in the analytic theory.  The results from 3D models are also
close to the results from the XZ models in Paper~I ($\etath=0.9$ and
$\etaturb=3.0$ at $\SigSFR=\SigSFRsun$).  In Equation (\ref{eq:etathnum}), the
control parameter for heating efficiency $\frad$ enters naturally into the
thermal yield coefficient $\etath$ since higher/lower heating efficiency (or
lower/higher shielding of FUV radiation) converts more/less radiation energy
from star formation feedback into thermal energy in the diffuse ISM.  Although
we have not directly explored variations in $\Psn/\Msn$ in the present
simulations (see SO12 for a study of this kind), the turbulent yield
coefficient $\etaturb$ would be expected to vary  proportional to the momentum
feedback per stellar mass $\Psn/\Msn$, which would introduce an additional
factor ${({\Psn/\Msn})/{3000\kms}}$ to the right-hand side of  Equation
(\ref{eq:etaturbnum}).  In addition to supernovae, other potential sources of
momentum injection associated with star formation include radiation forces and
cosmic rays (see OS11 for discussion and estimates).

The ratio of total-to-thermal pressure $\alpha$ can be obtained from equations
(\ref{eq:etathnum}) and (\ref{eq:etaturbnum}) as
\begin{equation}
\alpha =  1+\frac{\etaturb}{\etath}=
1+\frac{3.1}{\frad}
\left(\frac{\SigSFR}{10^{-3}\sfrunit}\right)^{0.03}.
\end{equation}
As discussed above, specific feedback momentum different from our chosen value
would introduce an additional factor $({\Psn/\Msn})/{3000\kms}$ in the second
term above.  For a fixed heating efficiency $\frad$, $\alpha$ is nearly
constant, as seen in Figure~\ref{fig:afw}(b). In the Solar neighborhood
($\frad=1$), $\alpha\approx 4$ is close to the fiducial value adopted by OML10
($\alpha=5$).

\subsection{Star Formation Scaling Relations}\label{sec:scaling}

\begin{figure*}
\epsscale{1.0}
\plottwo{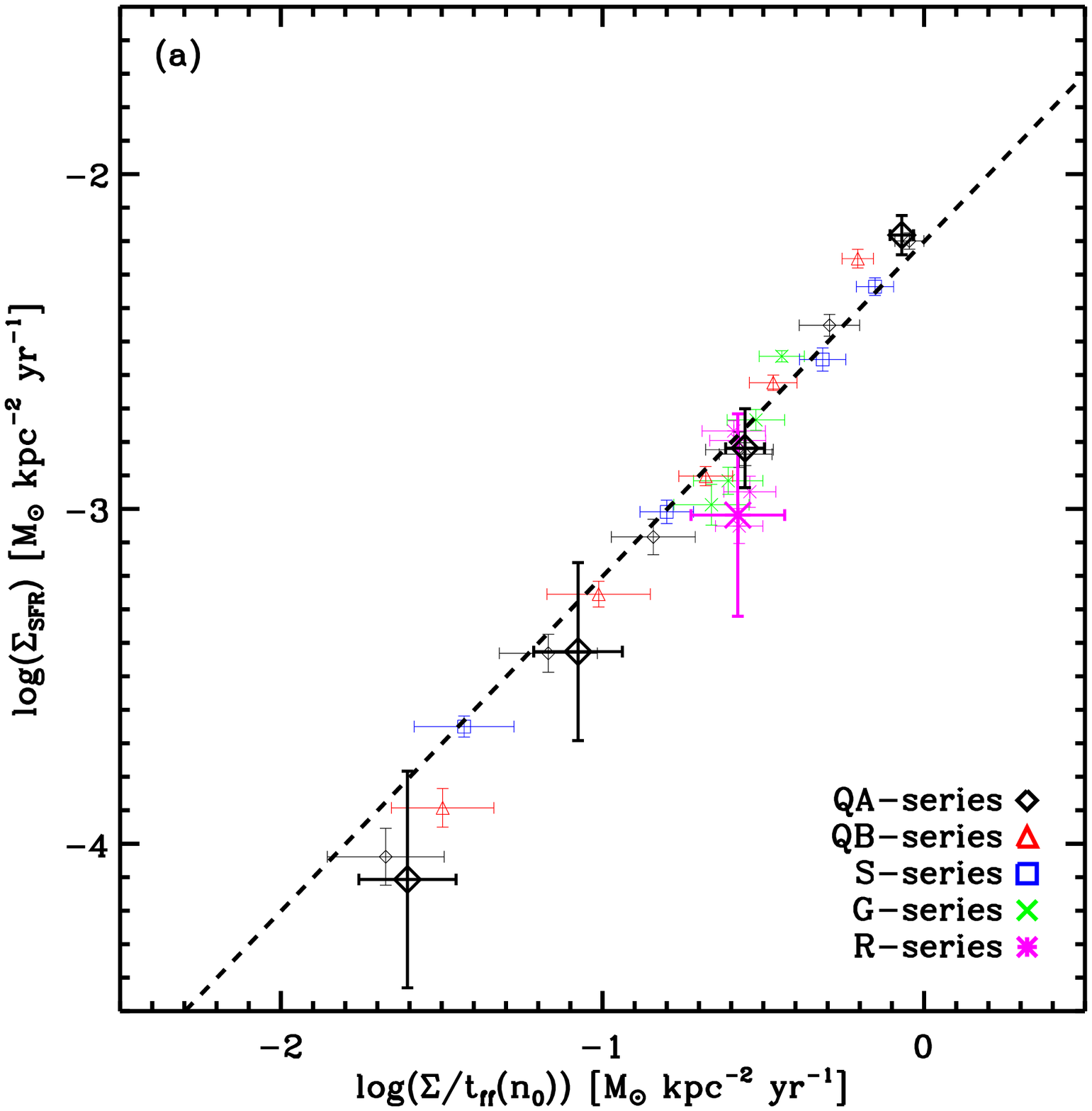}{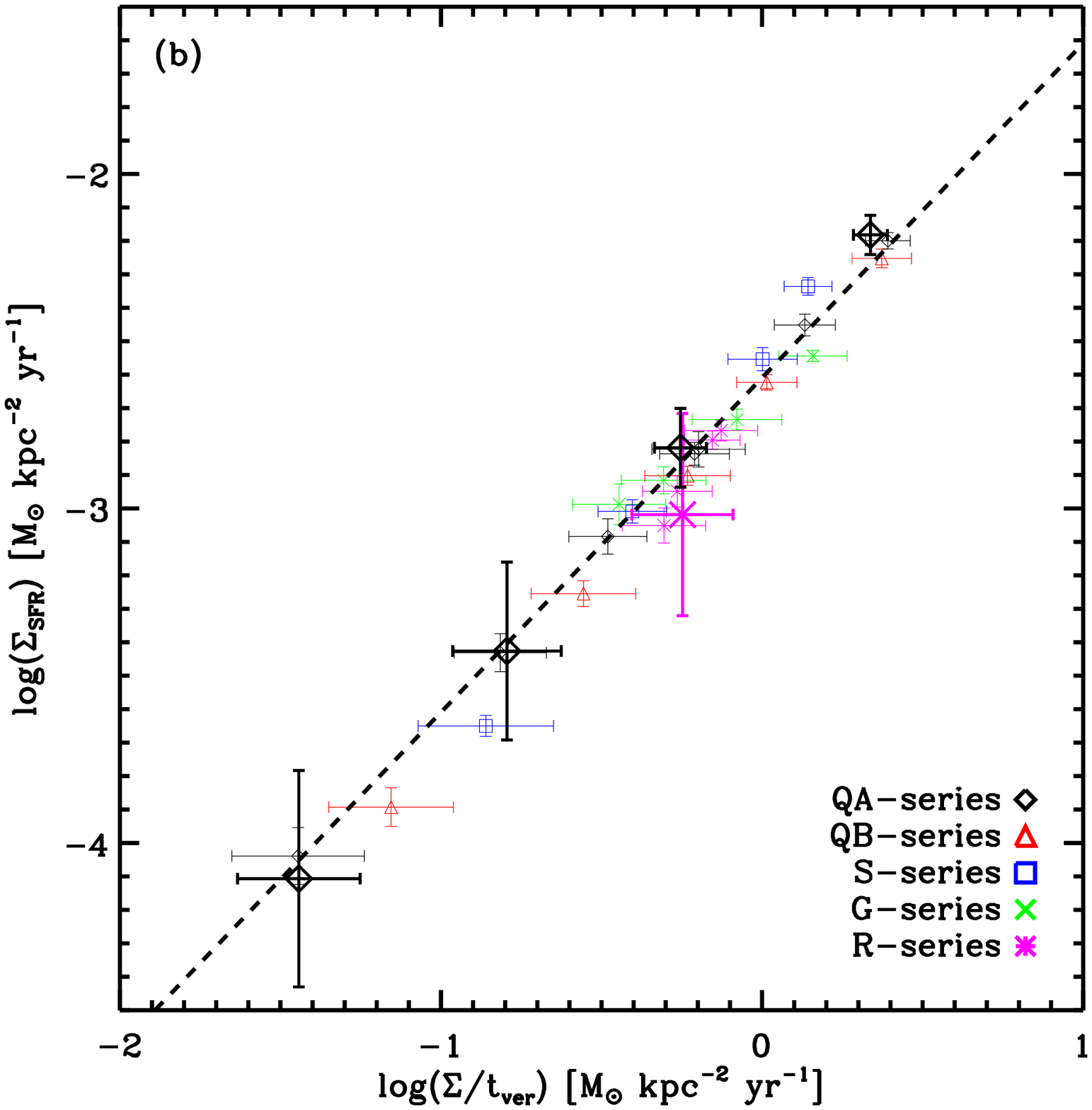}
\caption{ Measured SFR surface density $\SigSFR$ as a function of (a)
$\Sigma/\tff(\rhomid)$ and (b) $\Sigma/\tver$ for all 3D models.  The symbols
have the same meanings as in Figure~\ref{fig:v_sfr}.  The dashed lines in both
panels are the our best fits to all 3D models for an imposed unity slope, which
give the coefficients of $\eff(\rhomid)=0.006$ and $\ever=0.002$.
\label{fig:sfr_t}}
\end{figure*}

A common characterization of the SFR is in terms of a gas mass consumption
efficiency within a relevant dynamical time scale (e.g., \citealt{ler08} and
references therein):
\begin{equation}\label{eq:sfr_dyn}
\SigSFR=\edyn\frac{\Sigma}{\tdyn}
\end{equation}
A natural dynamical time to consider in disk galaxies is the free-fall time at
the mean midplane gas density (OS11), $\tff(\rho_0)=[3\pi/(32 G
\rho_0)]^{1/2}$, where $\rho_0=1.4m_p\rhomid$.  Using the time- and
horizontally-averaged values of $\rho_0$ measured in our simulations,
Figure~\ref{fig:sfr_t}(a) plots our measured values of $\SigSFR$ as a function
of $\Sigma/\tff(\rho_0)$ for all 3D models.  The best fit for an imposed linear
relation (dashed line) gives $\SigSFR \tff(\rho_0)/\Sigma  \equiv
\eff(\rhomid)=0.006$.  This value is close to typical values $\eff\sim 0.01$
inferred from observations of molecular gas \citep{kru07}, and similar to the
values for $\eff$ measured by SO12 in simulations of the starburst regime
($\SigSFR$ up to a few $\times\sfrunit$).

Formation of GBCs from diffuse gas is the first step to the star formation.  As
described in Section \ref{sec:evol}, in outer disk regions where the vertical
gravity from stars exceeds the (mean) vertical gravity from the gas, the
initial concentration of diffuse cold cloudlets to make GBCs is primarily
``falling'' toward the midplane of the stellar potential.  Under this
circumstance, the vertical dynamical time defined by
$\tver\equiv\Hdiff/\szdiff$ may be more relevant for gathering diffuse gas to
initiate star formation than the free-fall time.\footnote{ Using equation
(\ref{eq:PDE}), $\tver\equiv\Hdiff/\szdiff = \Sigdiff \szdiff/(\sqrt{2 \pi}
\Ptot)$ in dynamical equilibrium.  For diffuse-dominated regions, we therefore
expect $\tver \approx \left[(\pi^3/2)^{1/2} G \Sigma/\sz + (4 \pi G
\rhosd)^{1/2}  \right]^{-1}$.  In the limit that gas or stars  dominates the
gravity, this becomes $\tver \approx (\pi^2 G \rho_0)^{-1/2}$ or $\tver \approx
(4 \pi G \rhosd)^{-1/2}$, respectively.} Using the time- and
horizontally-averaged values of $\Hdiff$ and $\szdiff$ defined in Section
\ref{sec:stat}, in Figure~\ref{fig:sfr_t}(b) we plot the measured $\SigSFR$ as
a function of $\Sigma/\tver$ for all 3D models.  The dotted line denotes our
best fit for an imposed unity slope,  $\SigSFR \tver/\Sigma  \equiv
\ever=0.002$.  The free-fall time and the vertical dynamical time prescriptions
give rms fractional differences between measured and estimated $\SigSFR$ of
$24\%$ and $17\%$, respectively.

\begin{figure*}
\epsscale{1.0}
\plottwo{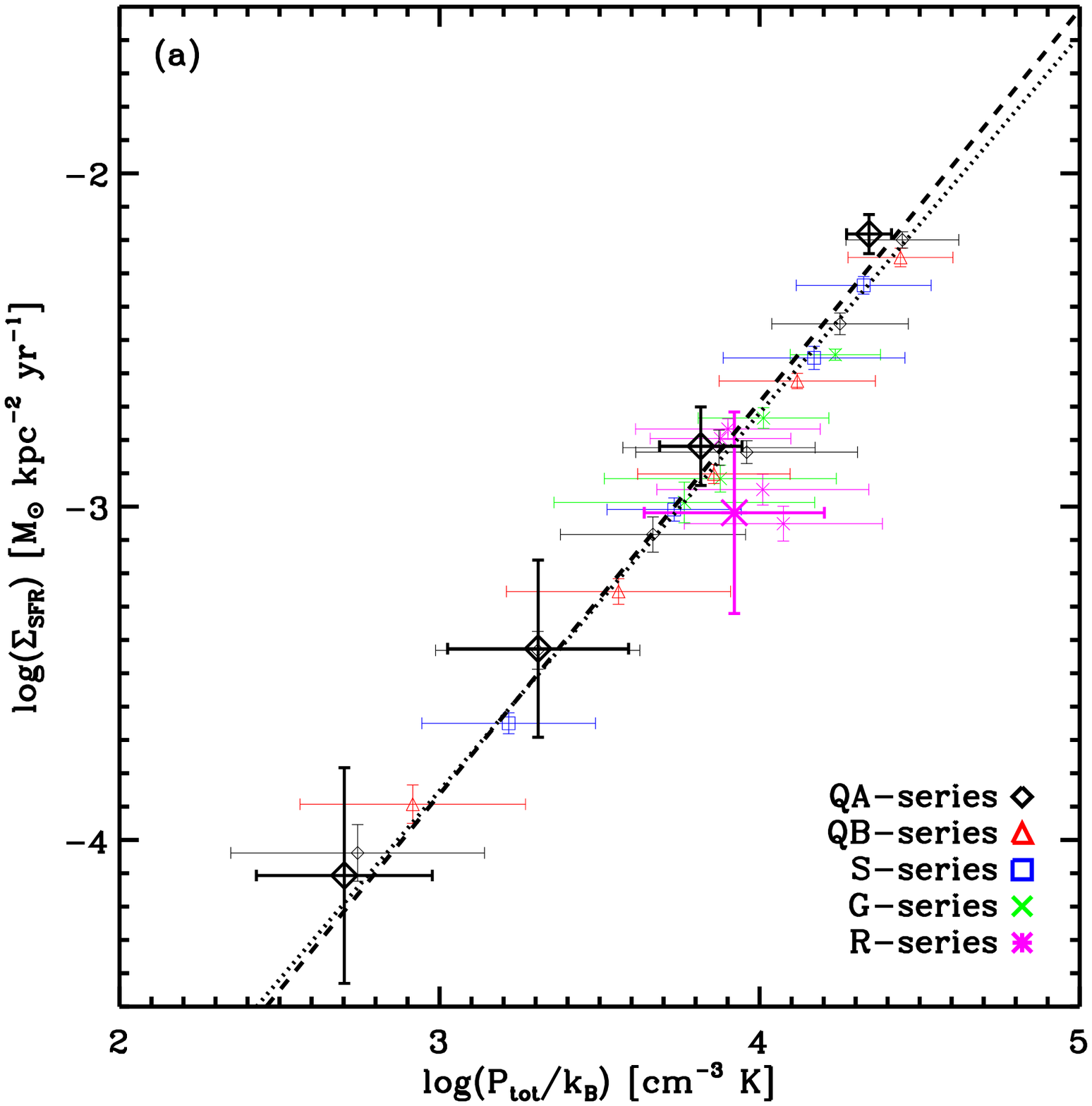}{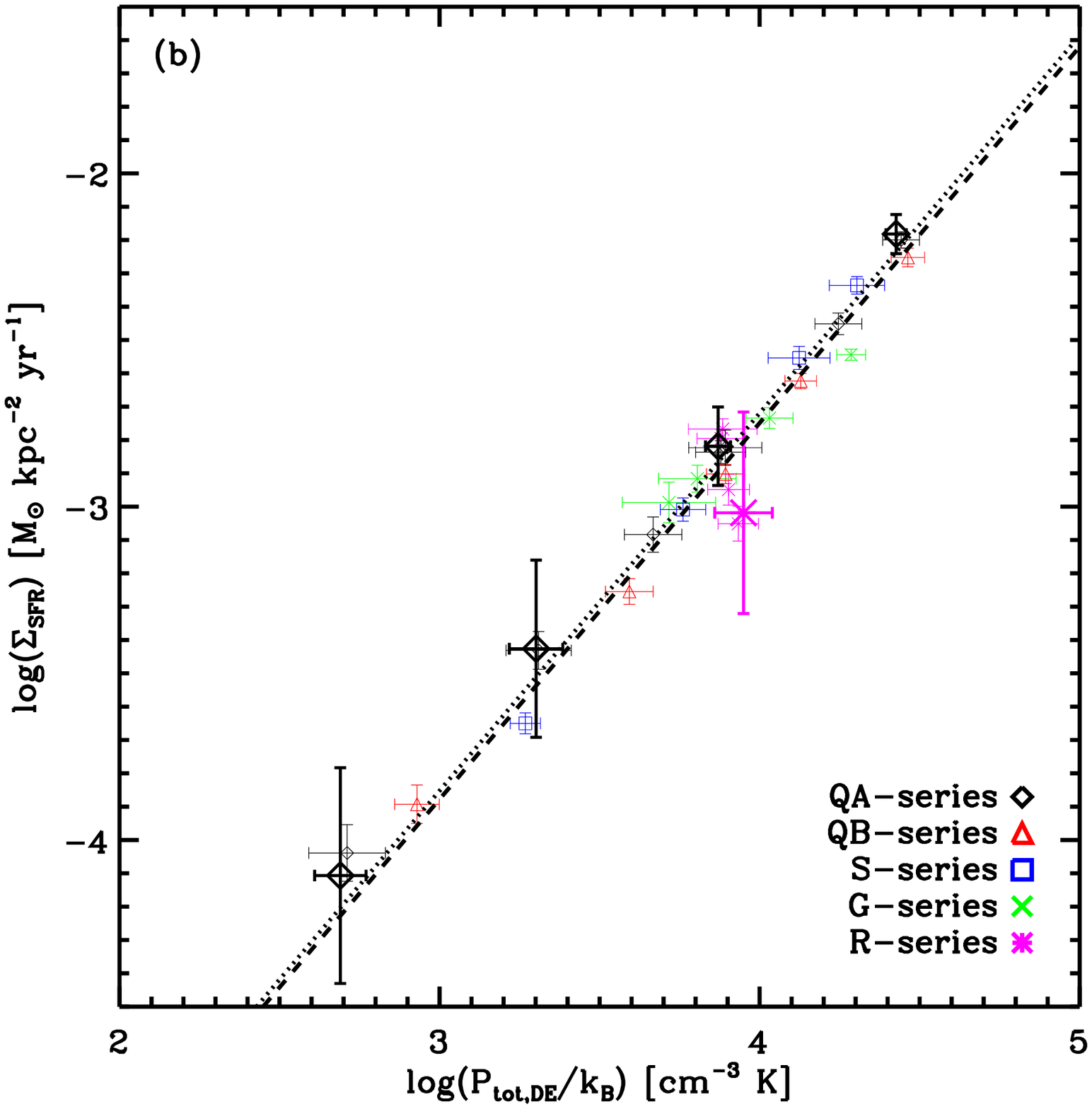}
\caption{ Measured SFR surface density $\SigSFR$ as a function of (a) the
measured midplane total pressure $\Ptot$ and (b) the predicted midplane total
pressure from vertical dynamical equilibrium $\PtotDE$.  The symbols have the
same meanings as in Figure~\ref{fig:v_sfr}.  In both panels, the dashed line
denotes our best fit (equation (\ref{eq:sfrp}) in [a] and equation
(\ref{eq:sfrpDE}) in [b]), and the dotted line plots the predicted thermal and
turbulent pressures using equations (\ref{eq:etath}) and (\ref{eq:etaturb})
with numerically calibrated yield coefficients $\etath$ (using $\frad=1$) and
$\etaturb$ from equations (\ref{eq:etathnum}) and (\ref{eq:etaturbnum}).  The
higher heating efficiency $\frad$ in the R series results in lower $\SigSFR$
for the same $\Ptot$ by increasing $\etath$ and hence $\eta$.
\label{fig:sfr_p} }
\end{figure*}

In the theory of self-regulated star formation (OML10, OS11, Paper~I), the
relationships among $\SigSFR$, $\Ptot$, and $\PtotDE$ are key.  Star formation
feedback replenishes the thermal and turbulent pressures, and the total
pressure supports the weight of the ISM.  Figure~\ref{fig:sfr_p}(a) shows
$\SigSFR$ as a function of $\Ptot$ for all 3D models.  The dotted line is
obtained from the sum of equations (\ref{eq:etath}) and (\ref{eq:etaturb}) with
the yield coefficients from equations (\ref{eq:etathnum}) and
(\ref{eq:etaturbnum}) for $\frad=1$, while the dashed line plots our best fit
omitting the R series:
\begin{equation}\label{eq:sfrp}
\SigSFR=2.1\times10^{-3}\sfrunit\rbrackets{\frac{\Ptot/\kbol}{10^4\pcc\Kel}}^{1.18}.
\end{equation}
Since both thermal and turbulent pressures separately satisfy nearly linear
relationships with $\SigSFR$ as shown in Section \ref{sec:equil}, the dotted
and dashed lines agree very well with each other.  Note that the
proportionality constant in the above relation for $\SigSFR$ vs $\Ptot$ is
roughly $4(p_*/m_*)^{-1}$, showing that the specific momentum injected by SNe
determines the coefficient of the $\SigSFR$--$\Ptot$ relation.

From equations (\ref{eq:etathnum}) and (\ref{eq:etaturbnum}), we obtain the
total feedback yield $\eta\equiv\etath+\etaturb$, which includes the effect of
$\frad$.  In regions of higher heating efficiency (or lower shielding), the
thermal feedback yield (and hence total feedback yield) increases.  The points
from the R series in Figure~\ref{fig:sfr_p} (magenta asterisks) thus spread
vertically with respect to model QA10.

In addition to the near-linear relationship between $\SigSFR$ and $\Ptot$
arising from the balance between energy gains and losses, the separate relation
$\Ptot=\PtotDE$ holds due to vertical dynamical equilibrium (Section
\ref{sec:equil}).  We thus can obtain a similar relation between $\SigSFR$ and
$\PtotDE$, as shown in Figure~\ref{fig:sfr_p}(b).  Our best fit analogous to
equation (\ref{eq:sfrp}) is:
\begin{equation}\label{eq:sfrpDE}
\SigSFR=1.8\times10^{-3}\sfrunit\rbrackets{\frac{\PtotDE/\kbol}{10^4\pcc\Kel}}^{1.13}.
\end{equation}
This relation provides a prediction for $\SigSFR$ in simultaneous thermal and
dynamical equilibrium, based only (via Equation (\ref{eq:PDE}) with $\fdiff
\sim 1$) on the total gas surface density and stellar/dark matter density
present at a given location in a galactic disk.

We remark that the ``dynamical time'' formulation of the SFR is in fact related
to the thermal/dynamical equilibrium formulation (based on momentum flux
matching), but the latter is more fundamental.  Considering the
turbulence-dominated case, balancing momentum flux requires $\rho_0 \vzdiff^2
\approx (1/4) (\Psn/m_*) \SigSFR $.  Using $\rho_0 \vzdiff^2 = \vzdiff \Sigdiff
\times \vzdiff/(\sqrt{2\pi} \Hdiff)$, this can be interpreted physically as the
statement that the dissipation of momentum on a vertical crossing time $\sim
\vzdiff \Sigdiff \times (\vzdiff/H)$ must be balanced by injection of fresh
momentum by star formation $\sim (\Psn/m_*) \SigSFR$.  The momentum balance
formula may be re-expressed as $\SigSFR (\Sigdiff/\tver)^{-1} \equiv
\epsilon_{\rm ver}= \vzdiff [(1/4) f_p \sqrt{2\pi} \Psn/m_*]^{-1}$.  A
relationship of the form in equation (\ref{eq:sfr_dyn}) for $\tdyn=\tver$ then
holds provided that $\vzdiff$ is constant.  From this point of view, the low
rate of ``gas consumption'' relative to the vertical dynamical time reflects
the ``inefficiency'' of converting ordered feedback momentum injected on a
small scale to a pervasive turbulent velocity dispersion on scales comparable
to the disk thickness.  That is, $\ever$ (or $\eff$) is small because $\vzdiff
[(1/4)\sqrt{2\pi}\Psn/m_*]^{-1}$ is small.  Alternatively, the kinetic energy
injected to the ISM by each feedback event, $\sim p_* \vzdiff$, is large
compared to the kinetic energy lost from the ISM, 
$\sim \vzdiff^2 m_*$, when gas is
locked up in stars leading up to the feedback event.  In contrast, the
efficiency of {\it momentum} replenishment per dynamical time is order-unity.

In conclusion, we find that a nearly linear relationship between $\SigSFR$ and
$\PtotDE$ is strongly supported by our numerical simulations.  Depending on
whether stars or gas dominate the local vertical potential well, this would
lead to either a relation $\SigSFR \propto \Sigma \sqrt{\rhosd}$ (typically
outer disks, as modeled here and in Paper~I) or $\SigSFR \propto \Sigma^2$
(typically starburst regions, as modeled by OS11 and SO12).  An increase in
either the heating efficiency or the turbulent driving efficiency would tend to
lower $\SigSFR$ for given values of the gas and stellar disk parameters
$\Sigma$ and $\rhosd$.

\section{Summary and Discussion}\label{sec:summary}

In this paper, we have carried out full three-dimensional simulations of the
turbulent, multiphase ISM in galactic disks, including the physical effects
from galactic differential rotation, gaseous self-gravity, vertical gravity due
to stars and dark matter, and cooling and heating as appropriate for atomic
gas.  We also incorporate thermal and kinetic feedback tied to recent star
formation, via a time-dependent heating rate coefficient and turbulent driving
from expanding supernova remnants.

We use our simulations to investigate the relationships between ISM properties
and star formation rates, and we also compare our numerical results to the
predictions of a recent theory for thermal/dynamical equilibrium of the ISM and
self-regulated star formation (OML10, OS11).  The theory posits that the rate
of star formation adjusts until the total energy (in turbulent driving and
thermal heating) supplied to the ISM by star formation feedback matches the
demand imposed by ISM losses (turbulence dissipation and interstellar cooling).
Force balance must also hold: the momentum flux (thermal and turbulent)
provided by feedback must match the vertical weight of the ISM in the total
gravitational field (gas, stellar, and dark matter gravity).  Our simulations
provide quantitative evidence in support of the theory.  This confirms our
previous results (Paper~I) based on radial-vertical simulations in the same
atomic-dominated regime, together with the simulations of SO12 for the
molecular-dominated starburst regime.

Our main findings are summarized as follows:

1. \emph{Vertical support of the disk.} -- Despite large amplitude temporal
fluctuations, we find that the mean total (thermal \emph{plus} turbulent)
midplane pressure $\Ptot$ matches the mean vertical weight of the gas
($\PtotDE$, as defined in equation~(\ref{eq:PDE})) within $12\%$.  Vertical
dynamical equilibrium in the highly turbulent, multiphase ISM has also
previously been demonstrated in simulations with a range of sources of
turbulence: expansion of \ion{H}{2} regions \citep{koy09b}, magnetorotational
instability \citep{pio07}, galactic spiral shocks \citep{kko10}, and SN
feedback (Paper~I; SO12; \citealt{hil12}).  These numerical results support the
assumption of effective ``hydrostatic'' equilibrium that is often utilized in
observations to obtain an estimate of the midplane pressure
\citep{won02,bli04,bli06,ler08}.  Equation (\ref{eq:PDE}) provides a general
expression for the pressure under vertical dynamical equilibrium.  When the
vertical gravity is dominated by stars and dark matter, as in the outer regions
of disk galaxies or dwarf galaxies, equation~(\ref{eq:PDEapprox}) gives
$\PtotDE\propto \Sigma\sqrt{\rhosd}$, while $\PtotDE\propto\Sigma^2$ if gaseous
self-gravity dominates -- typically in starbursts (OS11; SO12).

2. \emph{Thermal and turbulent energy replenishment from star formation
feedback.} -- Since turbulence decays within a flow crossing time \citep{sto98,
mac98} (comparable to the vertical dynamical time $\tver = H/v_z$ for the
neutral ISM disk) and the cooling time of the atomic gas is even shorter than
this dynamical time, both turbulent driving and heating are required to
maintain the quasi-steady state over a few orbital times that appears to hold
in normal disk galaxies.  Without continuous turbulent driving and heating, the
gas disk would rapidly collapse, leading to unrealistically high $\SigSFR$
\citep[e.g.,][]{dob11,hop11}.  In our simulations, star formation feedback
modeling SN explosions and photoelectric heating by FUV radiation replenishes
the turbulent and thermal energies  at a rate $\propto \SigSFR$.  For a given
gas surface density $\Sigma$, the cooling and turbulent dissipation rates per
unit mass increase when the gas disk's scale height decreases, because the
volume density $\rho$ increases.  However, higher $\rho$ also leads to more
rapid gravitational collapse and an increase in $\SigSFR/\Sigma$.  The star
formation rate can therefore adjust to meet the ISM's demands for energy
inputs.

In the OML10 theory, the demand to maintain thermal balance in multiphase
atomic ISM translates to the requirement that the thermal pressure at the
midplane $\Pth$ should lie between the maximum pressure of the warm phase
$\Pmax$ and the minimum pressure of the cold phase $\Pmin$
\citep{fie69,wol95,wol03}.  Here, we have compared the measured $\Pth$ to the
geometric mean pressure $\Ptwo\equiv({\Pmin\Pmax})^{1/2} \propto \SigSFR$,
showing that $\Pth\approx\Ptwo$ indeed holds with weak decreasing trend towards
increasing $\SigSFR$. The variation of $\Pth/\Ptwo$ is only $\sim 0.4{\;\rm
dex}$ for two orders of magnitude change in $\SigSFR$ and the measured $\Pth$.
This confirms both the hypothesis of OML10 and our previous numerical results
from Paper~I.

In OS11 (see also Paper~I and SO12), it is shown that the demand to
offset turbulent dissipation with turbulent driving from star
formation feedback translates to the requirement that $\Pturb =\rho
v_z^2 \sim \Sigma v_z /\tver$ is comparable to $\Pdriv \equiv
(1/4)(\Psn/\Msn)\SigSFR$.  Here, we have compared the measured
$\Pturb$ to $\Pdriv$, showing that $\Pturb\approx \Pdriv$ holds,
again with a weak decreasing trend as $\SigSFR$ increases.  Similar
to $\Pth/\Ptwo$, $\Pturb/\Pdriv$ varies only over $\sim 0.3{\;\rm
dex}$ for two orders of magnitude change in $\SigSFR$ and the
measured $\Pturb$. This confirms the hypothesis of OS11 and our
previous findings from XZ simulations (Paper~I), as well as
numerical results of SO12 for the starburst regime, in which
turbulent pressure is completely dominant over other pressures.

3. \emph{Thermal and turbulent energy yields.} -- The equilibrium pressures
depend on a balance between losses and gains: $\Pturb/\tver \sim$ (turbulent
energy driving/volume/time) $\sim (v_z/H) \times$ (turbulent momentum
driving/area/time)  so that $\Pturb = f_p (1/4) (p_*/m_*) \SigSFR$ in
equilibrium;  and $\Pth/t_{\rm cool} \sim$ (thermal heating/volume/time) so
that $\Pth = \Gamma \kbol T/\Lambda(T) \propto \frad \SigSFR$ in equilibrium.
We quantify the results in terms of feedback yield parameters $\etath$ and
$\etaturb$, corresponding to the ratio $\Pth/\SigSFR$ and $\Pturb/\SigSFR$,
respectively, in suitable units (see equations~(\ref{eq:etath}) and
(\ref{eq:etaturb})).  We find that $\etath$ and $\etaturb$ are nearly constant
(see equations (\ref{eq:etathnum}) and (\ref{eq:etaturbnum})) in our models
with $\frad=1$.  Thus, even though $\SigSFR$ and the individual thermal and
turbulent pressures vary over two orders of magnitude,  the ratio of the
total-to-thermal pressure $\alpha= 1+ \etaturb/\etath \sim4-5$ is nearly
constant (see Figure~\ref{fig:afw}).  From the above, the ratio $\Pturb/\Pth$
is approximately equal to the product of $\tver/t_{\rm cool}$ and $\dot {\cal
E}_{\rm turb}/\dot {\cal E}_{\rm th}$.  Although the input rate of turbulent
energy $\dot {\cal E}_{\rm turb}$ is small compared to the input rate of
thermal energy $\dot {\cal E}_{\rm th}$ ($\sim 4\%$ for the Solar
neighborhood), $\tver$ is much larger than $t_{\rm cool}$ (by a factor $\sim
100$ for the Solar neighborhood), so that the resulting $\Pturb$ exceeds
$\Pth$.  Since the dependence $\tver/t_{\rm cool} \propto \Sigma$ is
compensated by $\dot {\cal E}_{\rm turb}/\dot {\cal E}_{\rm th} \propto
\Sigma^{-1}$, the ratio $\Pturb/\Pth$ can remain roughly the same over a large
range of radii in disks.

The feedback yield increases if either the parameters governing gains increase
or those governing losses decrease.  The heating rate in our models is $\propto
\frad\SigSFR$.  Because $\Gamma$ is expected to be proportional to dust
abundance for photoelectric heating and $\Lambda$ is expected to be
proportional to metal abundance for C or O cooling, these dependences would
roughly cancel in $\etath$ if dust and metals vary together. However, the
heating rate for a given $\SigSFR$ is expected to increase in low-$A_V$ regions
of galaxies where FUV radiation can propagate farther (see
OML10); we explore this effect via varying  $\frad$.  Our models show that
$\etath$ increases with $\frad$, consistent with expectations.  As explicitly
explored in SO12 and expected from our result $\Pturb/\Pdriv\sim1$, $\etaturb$
depends nearly linearly on $\Psn/\Msn$.  In the present simulations, the value
of $p_*/m_* = 3000\kms$ we adopt is based on the value of $p_*$ for a
radiative-stage spherical supernova remnant propagating into a uniform medium
\citep{cio88,1998ApJ...500..342B, tho98}.  Supernovae are expected to be the
dominant source of turbulence in the diffuse ISM in many cases (see
\citealt{mac04}; OS11), but $p_*$ may still vary and it is important to
calibrate this parameter via high-resolution simulations of a realistic
(cloudy) ISM.  Although $\etaturb$ exceeds $\etath$ for the parameter sets we
have explored, potentially the reverse situation could hold in some galactic
environments.

4. \emph{A link between SFR and disk properties.} -- Given basic disk
properties (the total gas surface density $\Sigma$ and stellar+dark matter
density $\rhosd$), a theory of large-scale star formation should be able to
provide a prediction for $\Sigma_{\rm SFR}$.  In the OML10+OS11 model, this
prediction is obtained by requiring that thermal and turbulent balance
equations hold, and also that vertical pressure/gravity balance holds.
Simultaneous solution of these three balance equations leads to the result that
$\SigSFR$ varies nearly linearly with $\PtotDE$ (as defined in equation
(\ref{eq:PDE}); $\fdiff \sim 1$ for the regime studied here).  We verify this
result with our simulations, as seen in Figure~\ref{fig:sfr_p} and Equation
(\ref{eq:sfrpDE}).  Depending on whether the gas gravity or the gravity of the
stars dominates in Equation~(\ref{eq:PDE}), this can lead to $\SigSFR \propto
\Sigma^2$ for starburst systems (OS11; SO12), or to $\SigSFR \propto
\Sigma\sqrt{\rhosd}$ for normal outer-disk regions (OML10; Paper~I).

In recent local observations of the $\SigSFR$ vs. $\Sigma$ relation, the
low-$\Sigma$ regime is characterized by large scatter of composite data sets
\citep{big08} and systematic differences in power law indices $p$ in $\SigSFR
\propto \Sigma^{1+p}$ between individual galaxies \citep{won02, ler08}.  Our
simulations suggest that this may, in part, be due to ``projection'' on the
$\SigSFR$ vs. $\Sigma$ plane that neglects variations in the gravity of the
stellar disk (which would lead to $\SigSFR \propto \sqrt{\rhosd}$).   In fact,
\citet{bli06} and \citet{ler08} have demonstrated that molecular content and
the star formation rate increase with increasing stellar density.  In Paper~I
(see Figure 13(a) there), we showed that $\SigSFR$ at a given $\Sigma$ moves up
or down when the ratio of external (stellar) gravity to gas gravity (controlled
by $1/s_0$) is varied.  Our current simulations show the same effect.  In the
particular case where stellar and gaseous $Q$ values are constant (QA and QB
series), $\sqrt{\rhosd} \propto \Sigma$, which would lead to a steep relation
$\SigSFR \propto \Sigma^2$.

We show in Section \ref{sec:scaling} that dynamical time prescriptions are also
good descriptions of our numerical results; we find
$\SigSFR=0.006\Sigma/\tff(\rho_0)$ and $\SigSFR=0.002\Sigma/\tver$.  However,
we argue that the relation between pressure and star formation rate is more
fundamental and direct than these dynamical time prescriptions.  The relation
$\SigSFR \propto \PtotDE$  requires that $p_*/m_*$ is approximately constant,
whereas $\SigSFR \propto \Sigma/\tver$ requires that $v_z (p_*/m_*)^{-1}$ is
approximately constant, and $\SigSFR \propto \Sigma/\tff$ requires that $v_z
(g_z/G\Sigma)^{1/2}(p_*/m_*)^{-1}$ is approximately constant.  Our simulations
(as well as others; see below) do find approximately constant turbulent
velocity dispersions $v_z$.  The family of simulations we conducted also
happens to have $g_z/G\Sigma$ constant because $g_z \sim ( G \rhosd)^{1/2} v_z
$ and we adopted $\rhosd \propto \Sigma^2$.  When gas dominates the vertical
gravity (as in starburst regions), $g_z \sim G \Sigma$, which combined with
constant $v_z$ results in $\SigSFR \propto \Sigma/\tff$ as shown in SO12;
however, this need not be the case for galaxies in general.

The present models focus on the ISM regime in which diffuse gas dominates 
-- i.e. outer disks at low gas surface density $\Sigma$.  Moving to smaller
radii and regions of higher surface density, observations
\citep[e.g.][]{big08,ler08} show that gas in gravitationally-bound molecular
clouds exceeds the diffuse atomic gas, and that $\SigSFR \propto \Sigma$ with
little dependence on $\rhosd$.  One interpretation of the transition from outer
disks to these mid-disk regions is that it represents an increase of $t_{\rm
dest}/t_{\rm form}$ from small to large values, where these timescales
represent formation and destruction timescales for gravitationally bound
clouds.  If formation and destruction are in balance, then we expect $\fdiff =
t_{\rm form}/(t_{\rm form} + t_{\rm dest})$.  Dynamical equilibrium between
pressure and gravity in the diffuse ISM requires $\eta \SigSFR \sim \fdiff
\Sigma g_z \sim \Sigma g_z t_{\rm form}/(t_{\rm form} + t_{\rm dest})$.  If
$t_{\rm form} \sim \tver$, then $g_z t_{\rm form} \sim v_z$.  If furthermore
$v_z$ and $t_{\rm dest}$ are roughly constant (and $t_{\rm form} \ll t_{\rm
dest}$), this would lead to $\fdiff \ll 1$ and $\SigSFR \propto \Sigma$.
Testing whether this or another interpretation explains mid-disk observations
will require detailed models of cloud destruction, carefully following feedback
processes throughout the lives of massive stars.

5. \emph{Velocity dispersion driven by star formation feedback.} -- Balance
between turbulent driving and dissipation leads to $\vzdiff = (\sqrt{2\pi}/4)
f_p (\Psn/\Msn)\SigSFR(\Sigma/\tver)^{-1}$.   If star formation scales as
$\SigSFR=\ever\Sigma/\tver$ (see \S\ref{sec:scaling}), this yields
$\vzdiff=0.63 f_p \ever(\Psn/\Msn)\sim 3.8-5.6\kms$ for $\ever=0.2\%$,
$f_p=1-1.5$, and $\Psn/\Msn=3000\kms$, insensitive to $\Sigma$, $\rhosd$, and
$\SigSFR$.  The measured values of turbulent velocity dispersion in our
simulations are more or less constant over the whole range of parameters (see
Figure~\ref{fig:v_sfr}) since $f_p$ depends very weakly on $\SigSFR$ (see
Figure~\ref{fig:fp} and equation~(\ref{eq:fp})).  Many other recent simulations
have also found nearly constant velocity dispersions with respect to the input
SFR (e.g. \citealt{dib06,she08,age09,jou09,dob11}, SO12).  \ion{H}{1} velocity
dispersions reported in observations of the Milky Way and nearby face-on
galaxies \citep{dic90,zee99,hei03,pet07,kal09} show comparable values and
insensitivity to the $\SigSFR$.  Variations in the turbulent velocity amplitude 
driven by feedback could arise, however, if $\Psn/\Msn$ differs in more 
extreme environments.

Although turbulence driven by feedback appears to be crucial in preventing
runaway star formation, other mechanisms can also help to drive ISM turbulence.
These mechanisms include unsteady galactic spiral shocks
\citep{kim06,kko06,kko10,dob06}, magnetorotational instability
\citep{kim03,pio05,pio07}, large-scale gravitational instabilities
\citep{wad02,kim07,age09,bou10}, and cosmic inflow \citep{2010A&A...520A..17K}.
Large-scale gravitational instabilities may be particularly important in
driving non-circular motions in ULIRGs and high-redshift galaxies, where
measured velocity dispersions appear larger than in local disks   \citep[see
e.g. Fig. 14 of][]{2011ApJ...733..101G}.  However, it is important to keep in
mind that force balance in the vertical direction depends on velocity
dispersions at scales below the disk thickness (much smaller than has been
resolved in external galaxies), whereas gravitational instabilities primarily
drive turbulence at scales larger than the disk thickness.  Turbulence driven
by any of these mechanisms at scales smaller than the disk thickness would tend
to reduce the star formation rate, since it would partially offset the demand
for star formation feedback to match the required pressure.  Magnetic fields
also contribute pressure, but because the magnitude is smaller than turbulent
pressure and the magnetic scale height is large,  the fractional contribution
to offsetting the weight of the neutral ISM is small (OS11, \citealt{hil12}).
In future models, it will be interesting to quantify both how important various
sources (including diverse feedback processes) are for driving turbulent
velocity dispersions in the ISM, and the corresponding effects on limiting star
formation.

\acknowledgements{
This work was made possible by the facilities of the Shared Hierarchical
Academic Research Computing Network (SHARCNET:www.sharcnet.ca) and
Compute/Calcul Canada.  Part of numerical simulations were performed by using a
high performance computing cluster in the Korea Astronomy and Space Science
Institute. The work of C.-G.~K. is supported in part by a CITA National
Fellowship.  The work of E.~C.~O. was supported by grant AST0908185 from the
National Science Foundation.  The work of W.-T.~K. was supported by the
National Research Foundation of Korea (NRF) grant funded by the Korean
government (MEST), No. 2010-0000712.  }

\end{document}